\documentclass[aps,prb,twocolumn,showpacs,amsmath,amssymb,floats,showkeys,preprintnumbers,superscriptaddress]{revtex4-1}
\usepackage{graphicx}
\usepackage{color}
\usepackage{float}
\usepackage{ulem}
\usepackage{multirow}
\usepackage{dcolumn}
\usepackage{hyperref}
\usepackage[usenames,dvipsnames,svgnames,table]{xcolor}

\newcommand\T{\rule{0pt}{2.6ex}}       
\begin{document}
 
\title{High pressure layered structure of carbon disulfide}

\author{S.~Shahab Naghavi}

\affiliation{International  School for  Advanced Studies  (SISSA), and
CNR-IOM  Democritos  National  Simulation  Center,  Via  Bonomea  265,
I-34136 Trieste, Italy}
\affiliation{Department of Materials Science and Engineering, Northwestern University, Evanston, Illinois 60208, USA}

\author{Yanier Crespo}
\affiliation{International  Centre  for  Theoretical  Physics  (ICTP),
Strada Costiera 11, I-34151 Trieste, Italy }
\affiliation{International Institute of Physics, Av. Odilon Gomes de Lima, 1722, 
Capim Macio, CEP 59078-400, Natal RN Brazil}

\author{Roman Marto\v{n}\'{a}k}
\affiliation{Department of Experimental  Physics, Comenius University in Bratislava,
Mlynsk\'{a} Dolina F2, 842 48 Bratislava, Slovakia}

\author{Erio Tosatti}
\affiliation{International  School for  Advanced Studies  (SISSA), and
CNR-IOM  Democritos  National  Simulation  Center,  Via  Bonomea  265,
I-34136 Trieste, Italy}
\affiliation{International  Centre  for  Theoretical  Physics  (ICTP),
Strada Costiera 11, I-34151 Trieste, Italy }

\date{\today}
\begin{abstract} 
  Solid CS$_{2}$ is  superficially similar to CO$_{2}$, with  the same $Cmca$
  molecular crystal structure  at low pressures, which  has suggested similar
  phases  also  at  high  pressures.   We  carried  out  an  extensive  first
  principles evolutionary  search in order  to identify the  zero temperature
  lowest  enthalpy  structures of  CS$_{2}$  for  increasing pressure  up  to
  200\,GPa.  Surprisingly,  the molecular $Cmca$  phase does not  evolve into
  $\beta$-cristobalite as in CO$_{2}$, but transforms instead into phases HP2
  and  HP1, both  recently  described  in high  pressure  SiS$_{2}$.  HP1  in
  particular, with a wide stability  range, is a layered $P2_{1}/c$ structure
  characterized by  pairs of edge-sharing tetrahedra,  and theoretically more
  robust  than all  other CS$_{2}$  phases discussed  so far.   Its predicted
  Raman spectrum and  pair correlation function agree  with experiment better
  than those  of $\beta$-cristobalite, and further  differences are predicted
  between  their respective  IR spectra.   The  band gap  of HP1-CS$_{2}$  is
  calculated to  close under pressure yielding  an insulator-metal transition
  near  50 GPa  in agreement  with experimental  observations.  However,  the
  metallic density of states remains modest above this pressure, suggesting a
  different origin for the reported superconductivity.
\end{abstract}

\maketitle
\section{introduction}
\label{SEC:INTRO}  

The crystal  structure of  even extremely  stable molecular  carbon compounds
like CO$_{2}$ is known to transform  radically at high pressures where, above
$\sim$50\,GPa,     the     carbon     coordination     is     found,     both
experimentally~\cite{Iota1999}  and theoretically~\cite{Serra1999}  to switch
from   two   in   the   molecular   structure   $Cmca$   to   four   in   the
$\beta$-cristobalite    structure.~\cite{Santoro2012,     sun2009}    Similar
pressure-induced  structural transformations  can reasonably  be expected  to
occur in  a compound  such as  CS$_{2}$, which  is the  focus of  the present
study, and  which presents obvious  similarities to CO$_{2}$  Indeed CS$_{2}$
adopts  at zero  pressure  the  same $Cmca$  molecular  crystal structure  as
CO$_{2}$ at moderate  pressures between 15 and 50 GPa.   However, the binding
in CS$_{2}$ is  much weaker than in CO$_{2}$, with  a smaller electronic band
gap and a  positive formation enthalpy of about 88.7\,kJ/mol  as opposed to a
large  negative one  of $-393.509$\,kJ/mole  for CO$_{2}$.~\cite{CO2-Hf_1,Hf}
Thermodynamically,  that  makes  decomposition   and  phase  separation  into
elementary  carbon  and sulfur  a  thermodynamic  necessity for  CS$_{2}$  at
equilibrium already at ambient pressure and presumably even more so at higher
pressures.  In  spite of that intrinsic  thermodynamical metastability, solid
CS$_{2}$ phases  do exist, clearly for  kinetic reasons, and are  reported at
ambient  pressure  not to  decompose  in  measurable  times, at  least  below
$\approx$\,560\,K,~\cite{Dias2011,Bolduan1986}  a   temperature  rising  even
further at high pressures, possibly up to 1000\,K at 70\,GPa.~\cite{Dias2011}
At high  pressures but low  temperatures, X-ray  data have shown  evidence of
structural  transitions of  CS$_{2}$ from  the molecular  phase to  polymeric
phases  with C--S  coordinations  rising from  two  (\textit{Cmca}) to  three
(CS$_{3}$)  to   four  (CS$_{4}$),\cite{Dias2011}  and  further.    With  the
exception of  the CS$_{3}$ phase  near 20\,GPa, the high  pressure structural
behavior  has been  postulated so  far  to be  similar to  that of  CO$_{2}$,
implying (not  unreasonably) that CS$_{4}$ could  be $\beta$-cristobalite. In
CS$_{2}$  a detailed  interpretation of  high pressure  experimental data  is
further complicated by a large amount of structural disorder, particularly in
the CS$_{4}$ phase where only broad rather than sharp Bragg peaks are present
in the diffraction pattern.~\cite{Dias2011}  With that rationalization of the
fourfold coordinated  state of CS$_{2}$  near 50\,GPa through  simple analogy
with  CO$_{2}$,  density  functional  theory (DFT)  studies  have  considered
$\beta$-cristobalite  ($I\overline{4}2d$, also  referred to  as Chalcopyrite)
and tridymite  ($P2_{1}2_{1}2_{1}$), for  comparison with  experimental data.
Lacking  sharp   diffraction  peaks  or  other   distinctive  features,  that
comparison nevertheless  appears somewhat  elusive.  The  tridymite structure
shows   slightly  better   agreement   with  experiment   but  represents   a
thermodynamically less  likely candidate than $\beta$-cristobalite  since its
calculated DFT enthalpy  is 0.3--0.4\,eV/molecule higher.~\cite{Dias2011} The
electronic structure of either  fourfold coordinated crystal structure agrees
with    the   observed    metalization   in    the   region    from   40\,GPa
upwards.~\cite{Dias2011,Dias2013}  Besides that,  in more  recent experiments
performed  on the  high  pressure metallic  phases  of CS$_{2}$,  resistivity
showed evidence of  superconductivity at 4--6\,K over a  broad pressure range
from 50 to 172\,GPa.~\cite{Dias2013}

From the theoretical point of view it is of course inadequate to merely trust
the analogy  between CS$_{2}$ and  CO$_{2}$ and extend  it even to  very high
pressures.  To expand somewhat our view we instead consider that CO$_{2}$ and
CS$_{2}$ are members of a broader family of IV--VI AB$_2$ compounds including
SiO$_{2}$,  silica, and  SiS$_{2}$, neither  of  which is  molecular at  zero
pressure. Silica is very well known for a number of tetrahedrally coordinated
polymorphs  including  $\alpha$-quartz, $\alpha$-  and  $\beta$-cristobalite,
tridymite,  coesite,  etc.   Much  less  studied  until  recently,  SiS$_{2}$
exhibits  totally different  phases starting  off  at zero  pressure with  an
orthorhombic  (so-called  NP)  crystal  structure  made  up  of  edge-sharing
carbon-centered tetrahedra (see Ref.\onlinecite{SiS2_2015}) Very recently the
high     pressure     phases     of     SiS$_{2}$     were     experimentally
characterized~\cite{SiS2_2015} and a whole sequence  of phases HP1, HP2, HP3,
were  described where  the tetrahedra  did not  disappear but  simply changed
their mutual connectivity.

\begin{figure}
\centering
\includegraphics[width=0.99\linewidth]{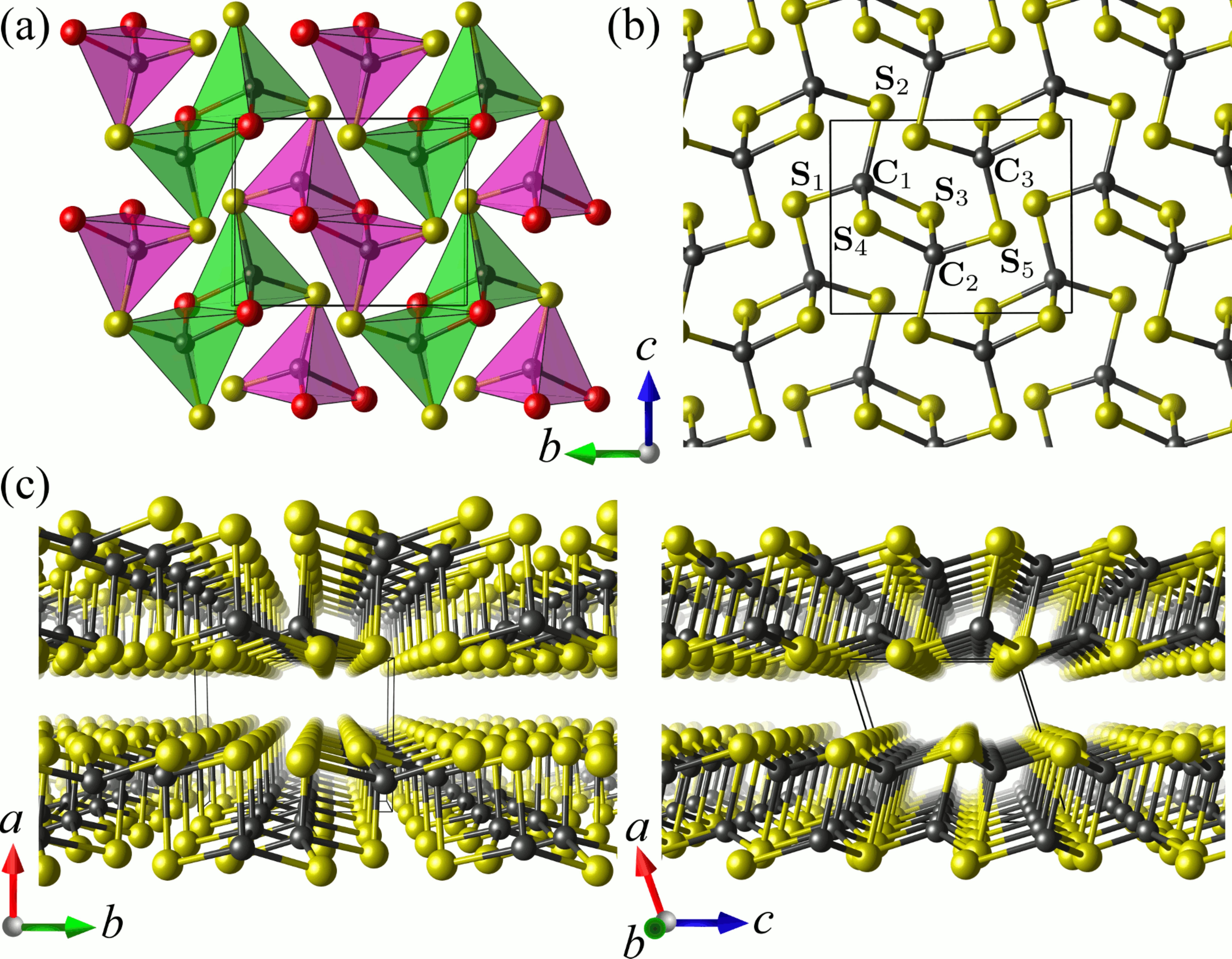}
\caption{  (Color online)  The  $P2_{1}/c$ layered  crystal structure,  named
{HP1} after that reported for  SiS$_{2}$,~\cite{SiS2_2015} viewed along $a$,
$b$  and  $c$-axes.   Each  layer  is  made  up  of  two  pairs  of  CS$_{4}$
tetrahedra. The two  tetrahedra in each pair  are joined by the  edge, with S
edge atoms  shown in red.   At $\approx$\,50\,GPa the interlayer  distance is
about 2.1\,{\AA}. Full structural data are in Table \ref{Tab:CIF}.}
\label{FIG:HP1}
\end{figure}

Based on the  limited available high-pressure data, solid CS$_{2}$  is at the
moment the least understood member of  this highly important class of solids.
Even  if starting  at zero  pressure  with the  same molecular  \textit{Cmca}
structure  which CO$_{2}$  adopts  at moderate  pressure,  CS$_{2}$ need  not
develop at very high pressures the same structures as CO$_{2}$.  To fill this
remaining  knowledge  gap  we   conducted  an  unbiased  theoretical  crystal
structure  search to  explore  high pressure  structures  of CS$_{2}$.   This
search revealed that at high pressures  where the carbon coordination is four
the $\beta$-cristobalite  structural motif is  indeed superseded by  a novel,
robust and unsuspected  layered network of tetrahedra.   Only while preparing
this  manuscript,   fresh  high  pressure  experimental   work  on  SiS$_{2}$
appeared,~\cite{SiS2_2015}  reporting  the  very same  structure,  designated
there  as  HP1.   Our  theoretical  search  now  found  the  tetrahedra-based
structure  of CS$_{2}$,  which  is lowest  in  enthalpy in  a  wide range  of
pressures  from  30  to  170\,GPa,  to be  structurally  identical  with  HP1
structure of  SiS$_{2}$~\cite{SiS2_2015} (see Fig. \ref{FIG:HP1}).   Based on
this  HP1 structure  of  CS$_2$  we calculated  the  Raman  spectra and  pair
distribution functions  and found it  to agree better with  experimental data
than those of  $\beta$-cristobalite, the high pressure  structure of CO$_{2}$
We also  obtained predictions for  the infrared (IR) absorption  spectra, not
yet  available experimentally,  that will  hopefully serve  in the  future to
experimentally  corroborate  or  discard  our  predictions.   The  calculated
electronic structure  of high  pressure metallized HP1-CS$_{2}$  GPa moreover
indicates a  small Fermi level  density of states, questioning  the intrinsic
nature of the observed superconductivity.

This paper  is organized as  follows. In Section \ref{SEC:METHOD}  we present
the details  of our  structural search and  subsequent analysis.   In Section
\ref{SEC:RESULTS}  we  present the  structural  results  of our  search.   In
Sections \ref{SEC:PHONONS}  and \ref{SEC:ELECTRONIC}  we analyze  the lattice
dynamics, Raman  spectra and electronic structure  of the new phase.   In the
final  section\ref{SEC:CONCLUSIONS}   we  summarize  the  results   and  draw
conclusions.

\section{Computational Methods}
\label{SEC:METHOD}

The  search  for low-enthalpy  structures  of  CS$_{2}$  was carried  out  by
exploiting  an  evolutionary algorithm  (EA),  as  implemented in  the  USPEX
code.~\cite{USPEX1,USPEX2,USPEX3,USPEX4,USPEX5M}   The   EA    was   run   in
conjunction  with  {\sl  ab-initio}  electronic  structure  calculations  and
relaxations  based   on  standard  density  functional   theory  (DFT).   The
Perdew-Burke-Ernzerhof (PBE)~\cite{PBE} generalized gradient approximation as
implemented in VASP ({\sl  Vienna Ab-initio Simulation Package})~\cite{VASP1}
employed  the projector-augmented  plane wave  (PAW) method.~\cite{PAW1,PAW2}
The energy cutoff for the plane-wave basis  was set to 550\,eV to ensure full
convergence.  The Brillouin Zone was  sampled by Monkhorst-Pack meshes with a
resolution  of $2\pi\times  0.05$\,{\AA}$^{-1}$. Since  at low  pressures the
dispersion forces  are important  we employed  for enthalpy  calculations the
optB86b-vdW scheme \cite{Klimes2011,Klimes2011_2} based  on the van der Waals
density  functional.\cite{Dion2004}   Lattice  zero-point energies  were  not
included  and temperature  was assumed  to be  zero throughout.   The phonon,
Raman  and  IR  spectra  were   instead  calculated  in  linear  response  as
implemented in  the {\sc Quantum ESPRESSO}  suite of programs within  the LDA
approximation.~\cite{pwscf} For  that, the GGA input  structures were further
relaxed with LDA before carrying out the phonon calculations.

Brute-force application of  the EA structural search algorithm  to a compound
such as CS$_{2}$ with a positive  formation enthalpy should lead by necessity
to chemical decomposition and outright  disappearance of the compound itself.
As was said above, even at ambient pressure molecular CS$_{2}$ is metastable,
meaning it is  locally stable mechanically, and temporarily  surviving due to
exceedingly  slow kinetics,  while  intrinsically  unstable on  thermodynamic
grounds.   The  EA  search  ignores  kinetics  and  will  therefore  lead  to
decomposition, given a large enough supercell.   In this study, where we wish
to find  and study the  high pressure  metastable phases of  the undecomposed
compound, a strict CS$_{2}$ stoichiometry was  assumed with an EA cell chosen
to contain  12 atoms, i.\,e.  four  CS$_{2}$ molecular units. Even  if small,
that cell  is actually  still large  enough to  show decomposition  and phase
separation into carbon and sulfur in an unrestricted EA search.  We therefore
artificially prevented decomposition  by constraining the EA  search to avoid
the formation of  C---C and S---S first-neighbor bonds.  This  allows to find
the high pressure  crystal structures where decomposition,  total or partial,
is artificially  suppressed.  It  should be noted  that while  the artificial
nature of this constraint is such that  it may endanger the formation of some
denser structures  at higher  pressures, still various  octahedral structures
were also created,  but their calculated enthalpies  were considerably higher
with respect  to the  tetrahedral ones.  Therefore,  we will  by construction
limit  ourselves to  explore metastable  structures with  coordination larger
than four.  With this constraint,  the crystal structure search was performed
at 0, 26, 38, 75, 120, and 170 GPa.  The EA initial population (the number of
structures in the starting generation) was  set to 120, a large number chosen
in order  to densely  sample the  configuration space  of the  random search,
later reduced to 30 in the following generations.

\begin{table}
  \caption{\label{tab1}  Structural data  of  the new  phases.  As seen  from
Wyckoff positions, the number of CS$_{2}$ formula units/cell is $Z=4$ for all
the phases except HP2 where $Z=12$.}

\begin{ruledtabular}
\begin{tabular}{p{0.25cm}ccccccc}
\multirow{8}{*}{\rotatebox{90}{P=50 GPa}} 
&\multicolumn{6}{c}{HP1---($P2_{1}/c$)} \\
& &  & $x$ & $y$ & $z$ & Wyck.  \\
& \multirow{3}{*}{Coordinate}
&   C  &  0.153 &  -0.355  & -0.322  &  4e \\
& & S  &  0.277 &  -0.091  & -0.435  &  4e \\
& & S  &  0.235 &   0.126  &  0.023  &  4e \\
& &    &        &          &         &     \\
& \multirow{2}{*}{Lattice}  
&   $a$  &  $b$ &   $c$  &  $\beta$  &  $Vol.$ \\ 
& & 4.86 & 5.59 & 4.74 & 110.02 & 120.95 \\
\hline \T
\multirow{14}{*}{\rotatebox{90}{P=20 GPa}} 
&\multicolumn{6}{c}{HP2---($P2_{1}/c$)} \\
& &  & $x$ & $y$ & $z$ & Wyck.  \\
&\multirow{9}{*}{Coordinate}
&   C &   0.355  & -0.133  & -0.499  &  4e \\ 
& & C &   0.213  & -0.365  & -0.332  &  4e \\ 
& & C &  -0.061  &  0.362  & -0.336  &  4e \\ 
& & S &  -0.438  & -0.105  &  0.412  &  4e \\ 
& & S &   0.067  & -0.131  &  0.423  &  4e \\ 
& & S &   0.415  & -0.374  & -0.419  &  4e \\ 
& & S &   0.275  & -0.133  & -0.243  &  4e \\ 
& & S &  -0.079  & -0.388  & -0.412  &  4e \\ 
& & S &   0.227  &  0.388  & -0.251  &  4e \\ 
& &    &        &          &         &     \\
&\multirow{2}{*}{\parbox{1.5cm}{Lattice}}
&   $a$  &  $b$ &   $c$  &  $\beta$  &  $Vol.$ \\ 
& & 6.21 & 6.42 & 11.49  & 105.01 & 442.17 \\
 \hline \T
\multirow{7}{*}{\rotatebox{90}{P=10 GPa}} 
&\multicolumn{6}{c}{NP---($Ibam$)} \\
& &  & $x$ & $y$ & $z$ & Wyck.  \\

&\multirow{2}{*}{Coordinate} 
&   C &   0.000  &  0.000  &  0.250  &  4a \\ 
& & S &   0.122  & -0.206  &  0.000  &  8j \\
& &    &        &          &         &     \\
&\multirow{2}{*}{\parbox{1.5cm}{Lattice}}  
&   $a$  &  $b$ &   $c$  & $\alpha,\beta,\gamma$  &  $Vol.$ \\ 
& & 6.96 & 5.09 & 4.83   &  90 &  171.21 \\
\end{tabular}
\end{ruledtabular}
\label{Tab:CIF}
\end{table}

\section{Crystal Structures, Pair Correlations and X-Ray Diffraction
  Spectra}
\label{SEC:RESULTS}

\begin{figure}[t!]
\centering
\includegraphics[width=0.99\linewidth]{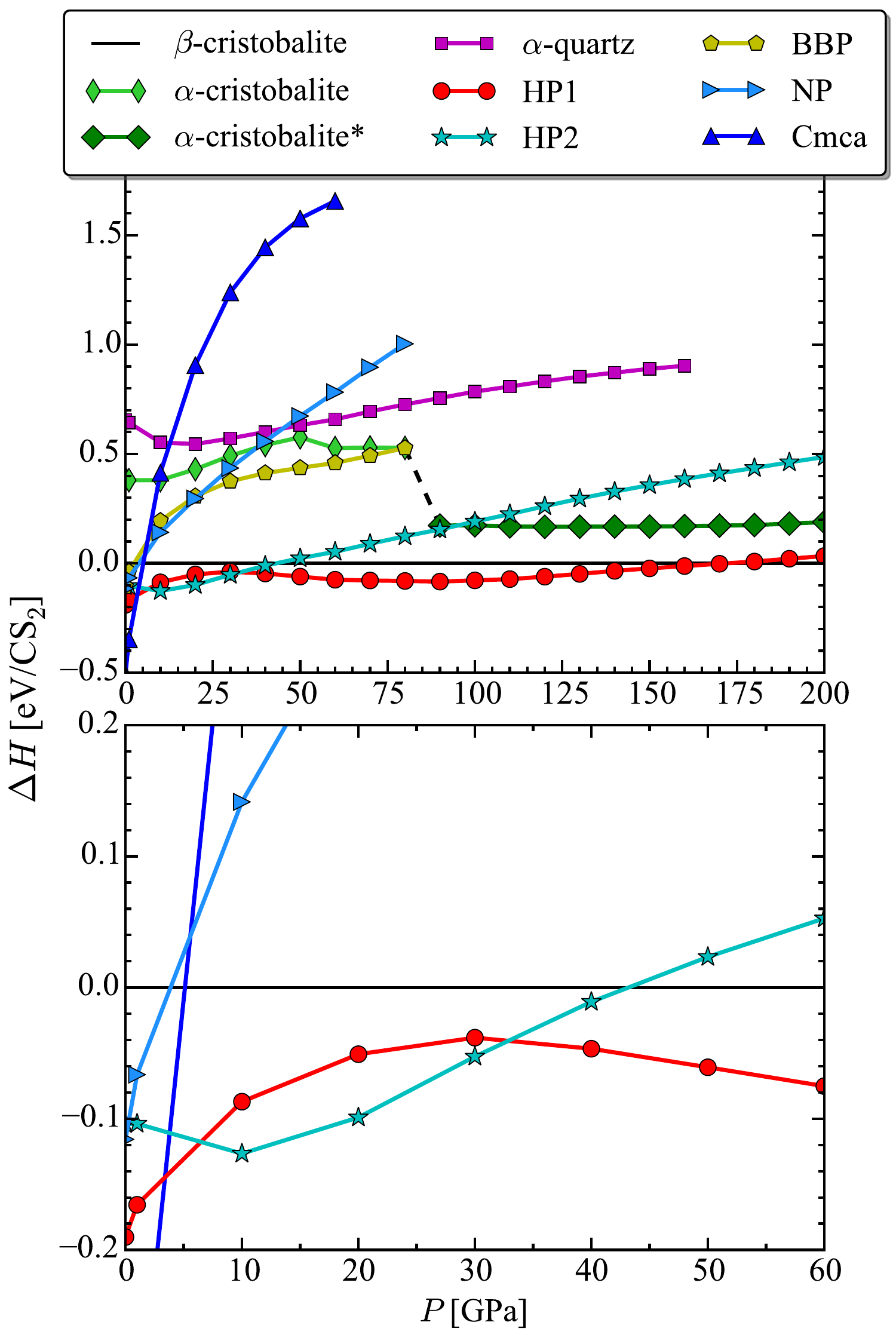}
\caption{(Color  online)   Zero  temperature  enthalpy  for   some  optimized
structures   of   CS$_{2}$,    plotted   relative   to   $\beta$-cristobalite
($I\overline{4}2d$).   No structures  with carbon  coordinations of  three or
larger  than  four are  considered.   Structures  HP1  ($P2_{1}/c$ ),  HP2  (
$P2_{1}/c$) and  NP ( \textit{Ibam}  ) are  based on CS$_4$  tetrahedra.  The
relative  stability  of HP2  between  10--30\,GPa  is marginal  and  possibly
unreliable.  $\alpha$-cristobalite* ($P2_{1}2_{1}2_{1}$ (No. 19). ) is a high
pressure version  of regular $\alpha$-cristobalite ($P4_{1}2_{1}2$  (No. 92))
where symmetry is  reduced through a rotation of  the tetrahedral environment
of carbon atom.}
\label{FIG:1}
\end{figure}

The EA  search produced a  large variety  of structures. In  particular, near
$P=0$  we reproduced  the known  molecular $Cmca$  structure, validating  the
scheme. Since  experimentally the CS$_{4}$ tetrahedral  structure was created
at  30\,GPa,~\cite{Dias2013} we  further  focused on  search for  tetrahedral
structures  at pressures  above  26\,GPa.  In  Fig.\ref{FIG:1}  we present  a
selection of  low-enthalpy phases,  that is molecular  $Cmca$, and  among the
non-molecular,   $\alpha$-  and   $\beta$-cristobalite,  compared   with  the
tetrahedra-based phases  HP1 and  HP2.  To complete  the comparison,  we also
added in  the $\alpha$-quartz  structure well known  from SiO$_2$.   While as
explained above  all these  structures are metastable  against decomposition,
those  which  we  will  describe  are  at  least  mechanically  stable  under
structural  optimization  and  and  dynamically stable  as  shown  by  phonon
calculations (see  e.g., Fig.~\ref{FIG:5}), thus representing  local enthalpy
minima.  At any  given pressure these phases compete, so  they are all doubly
metastable except for that with lowest  enthalpy.  It makes sense to consider
them all  anyway, as  that broader picture  clarifies the  relative stability
margins -- and also because  metastable phases commonly appear experimentally
for kinetic reasons.

Except for  0\,GPa where molecular $Cmca$  was found to prevail  in agreement
with experiment,  the most important structure  obtained by the EA  search at
high pressures was the $P2_{1}/c$ (HP1). This monoclinic layered structure is
quite  interesting. We  optimized  its  structure at  50\,GPa  and found  the
structural  parameters  as  listed   in  Table~\ref{Tab:CIF}.   As  shown  in
Fig.~\ref{FIG:HP1} the  carbon coordination is  four, but is  quite different
from $\beta$-cristobalite.  Each layer  consists of four CS$_{4}$ tetrahedra,
or more  accurately two pairs  of tetrahedra.  Tetrahedra of  different pairs
(different  colors in  Fig.~\ref{FIG:HP1})  share a  corner  sulfur, but  two
tetrahedra  in the  same pair  (same color)  share an  edge, made  up of  two
sulfurs. Simple  as it  looks, this  structure came  in at  first as  a total
surprise---we designated it as "shahabite", in  the lack of an existing name.
We  found   subsequently  that  the   very  same  structure  has   been  very
recently~\cite{SiS2_2015} observed  and called  HP1 in  the phase  diagram of
SiS$_{2}$,  at  the  much  lower  pressure  of  2.8\,GPa  (actually,  it  had
apparently been observed in SiS$_{2}$  long time ago\cite{Guseva1991} but not
resolved).

In Ref.\onlinecite{SiS2_2015}  another monoclinic phase with  the space group
$P2_{1}/c$ denoted as HP2 was observed  to follow HP1 at pressure of 3.5\,GPa
in  SiS$_{2}$. Unlike  the HP1,  which is  layered and  involves four  CS$_2$
units, the HP2 phase consists of 12 CS$_{2}$ units, arranged in a 3D covalent
network.  For the  sake of completeness we calculated  enthalpies in CS$_{2}$
for optimized HP2, and for the chain-like orthorhombic phase with space group
\textit{Ibam}  which was  denoted  in Ref.\onlinecite{SiS2_2015}  as NP,  the
structure of  SiS$_{2}$ at  ambient pressure.   The structural  parameters of
these  theoretical CS$_{2}$  phases (HP1,  HP2  and NP)  at various  relevant
pressures are listed in Table.~\ref{Tab:CIF}. Among phases not shown here and
not    further     discussed    is    tridymite,    also     considered    by
Ref.\onlinecite{Dias2011},  which  we  found   at  50\,GPa  to  be  unstable,
spontaneously converting into a low-symmetry structure.

\begin{figure}[!t]
\centering
\includegraphics[width=1.0\linewidth]{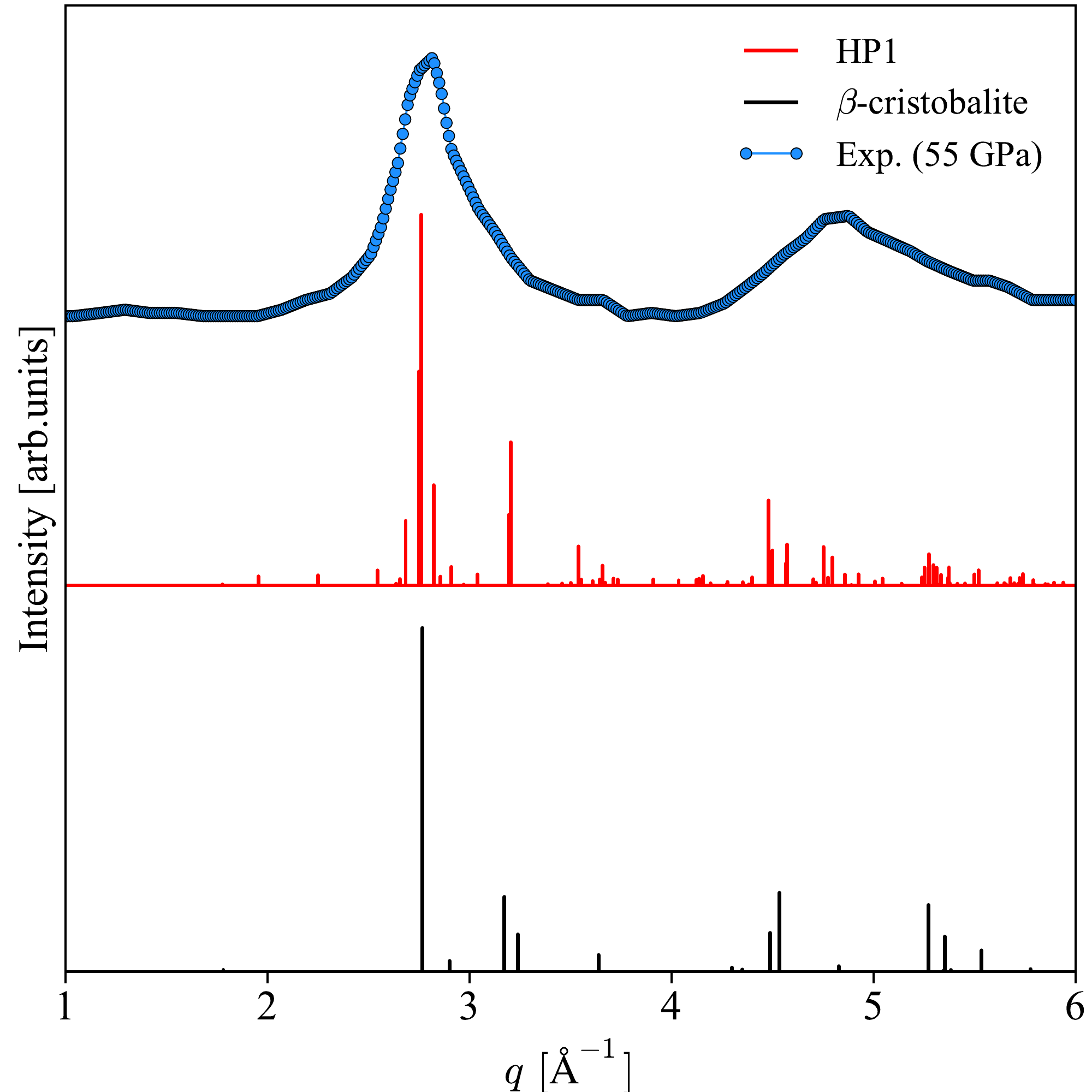}
\caption{Calculated   X-ray    diffraction   patterns   of   the    HP1   and
$\beta$-cristobalite structures  of CS$_{2}$ at P\,=\,55\,GPa,  compared with
experimental data of Ref.~\onlinecite{Dias2011}}
\label{fig-diff}
\end{figure}

The  relative  enthalpies  calculated  for  these  structures  are  shown  in
Fig.\ref{FIG:1}. The  $Cmca$ molecular phase  prevailing at low  pressures is
quickly  supplanted   by  $I\overline{4}2d$  $\beta$-cristobalite   at  about
$\approx$10\,GPa. At the same  time, however, the $P2_{1}/c$ tetrahedra-based
structures appear,  preempting this  transformation and replacing  the $Cmca$
molecular structure  already at about  5\,GPa. Upon increasing  pressure, the
two structures HP1  and HP2 remain nearly iso-enthalpic up  to 30\,GPa. Above
that  pressure  the HP1  layered  structure  clearly prevails,  its  enthalpy
remaining lower  than that of  $\beta$-cristobalite by a  substantial amount,
exceeding the estimated computational errors  of about 10 meV/molecular unit,
up until 160\,GPa---a regime where, however, carbon coordinations larger than
four will  take over.~\cite{Dias2011} The  same HP1 structure which  we found
here appears  in SiS$_{2}$  at much  lower pressures than  in CS$_{2}$  as is
natural given  the shorter  bond lengths and  smaller compressibility  of the
carbon compound.

It is  interesting to rationalize  the finding of edge-sharing  tetrahedra in
high pressure  CS$_{2}$, which  behaves similarly  to low  pressure SiS$_{2}$
whose  phases  consist of  variously  packed  tetrahedra.  The  stability  of
tetrahedra-based phases  in a  carbon compound  such as  CO$_2$ is  denied by
Pauling's third rule for ionic crystals which states that carbon edge-sharing
has a  destabilizing effect as it  brings the positive carbon  ions too close
together increasing their Coulomb repulsion energy.  The question is then why
this obstacle does not arise in  CS$_{2}$. In order to compare the importance
of this effect in CO$_{2}$ and CS$_{2}$ we structurally optimized CO$_{2}$ in
the HP1  structure at 50  GPa. In agreement with  Pauling, we found  for this
structure  a much  higher enthalpy  of 0.4\,eV/molecule  with respect  to the
stable phase  $\beta$-cristobalite, showing that edge-sharing  in CO$_{2}$ is
indeed unfavorable.  To confirm that  this is  due to ionicity  we calculated
Bader charges\cite{Bader1990,Henkelman2006} for  the C, S and O  atoms in the
HP1 structure. Strikingly whereas in CO$_{2}$ the partial charge of carbon is
about  $+$2,  it  turned out  to  be  about  $-$0.5  in CS$_{2}$.   The  bond
polarization in CS$_{2}$ is not only of small magnitude, but opposite to that
in CO$_{2}$. This finding explains why edge-sharing in CS$_{2}$ does not have
the destabilizing effect it has in CO$_{2}$.  A second significant difference
between  CO$_{2}$ and  CS$_{2}$ originates  from the  different chemistry  of
oxygen  and sulfur  which  becomes  relevant at  higher  pressures where  the
chalcogen  binds  two  carbons.  In  this  configuration,  oxygen  hybridizes
sp$^{3}$,  favoring  a bond  angle  of  about 109$^{\circ}$;  sulfur  instead
prefers $p$ orbital binding without hybridization, and a bond angle closer to
90$^{\circ}$.        (See      a       detailed      discussion       e.g.,in
Ref.\onlinecite{james_book}). Therefore,  it is  not surprising to  find that
CO$_{2}$ should  adopt the  $\beta$-cristobalite structure where  the C--O--C
bond angles  are 106$^{\circ}$ and  115$^{\circ}$ while CS$_{2}$  prefers the
edge sharing tetrahedra with C--S--C bond angle of 90$^{\circ}$.

\begin{figure}[t!]
\centering
\includegraphics[width=0.99\linewidth]{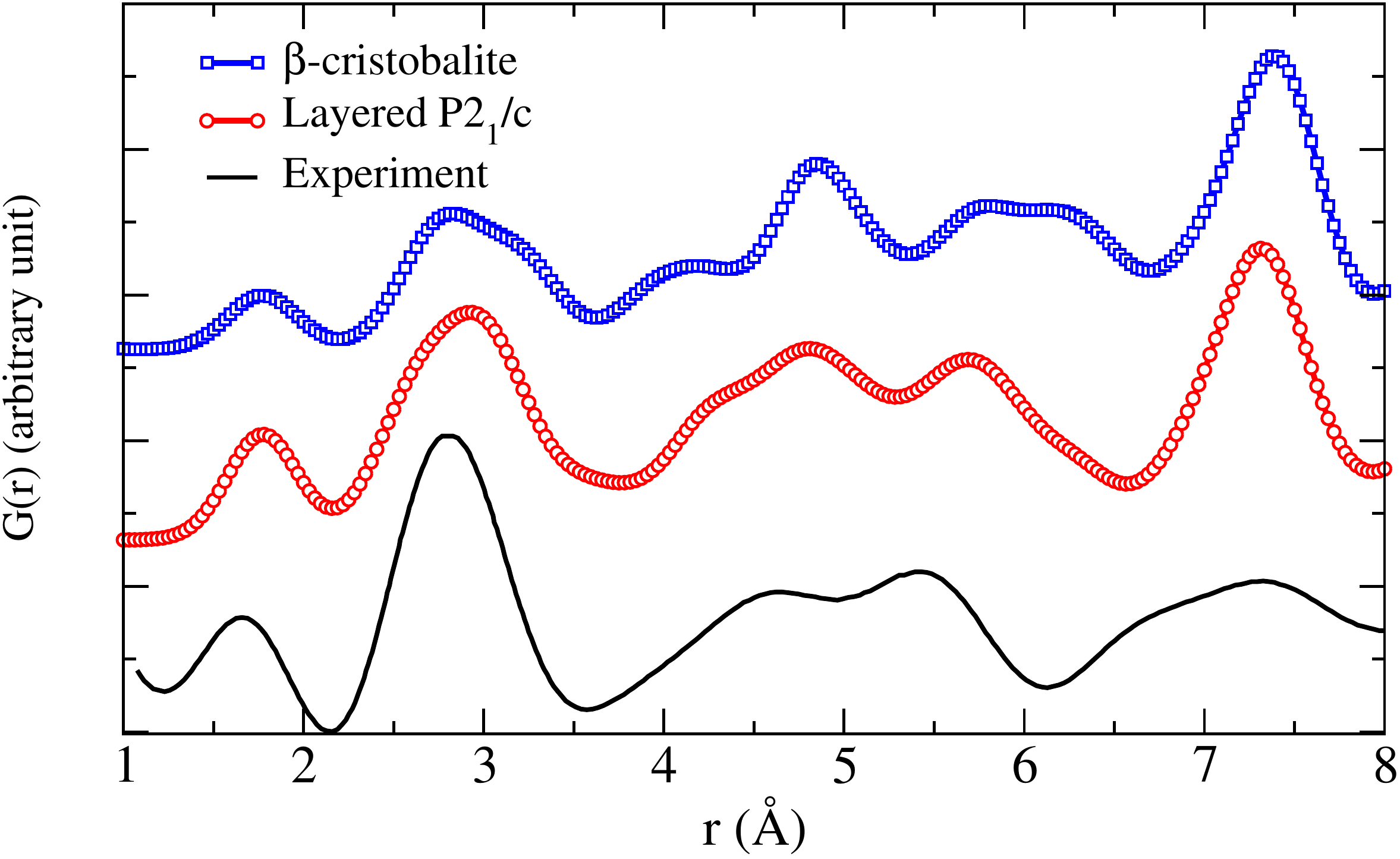}
\caption{(Color online)  Calculated pair  correlations $G(r)$ at  50\,GPa for
$\beta$-cristobalite  (green)  and  layered  HP1 (blue)  in  comparison  with
experimental  data  at  55\,GPa taken  from  Ref.~\onlinecite{Dias2013}.   An
empirical Gaussian  broadening was  applied to the  $G(R)$ of  the respective
perfect  crystals in  order to  mimic the  structural disorder.  The Gaussian
width was chosen  in order to get  for the C--S bond peak  at 1.75\,{\AA} the
same width as in experiment.\cite{Dias2011} }
\label{FIG:Gr}
\end{figure}

Fig.\ref{fig-diff} shows  the calculated diffraction  pattern of the  HP1 and
$\beta$-cristobalite structures of CS$_2$  at 50\,GPa. Comparing the position
of the Bragg  peaks with those of  broad peaks of the  X-ray structure factor
$S(Q)$ in experiment at 55\,GPa  (Fig.3 in Ref.\onlinecite{Dias2011}) we find
roughly  the same  agreement for  both crystalline  structures.  In  order to
discuss  and compare  more  realistically the  direct-space pair  correlation
function, we carried out an {\sl ab-initio} Molecular Dynamics simulation for
HP1-CS$_{2}$ and $\beta$-cristobalite CS$_{2}$  at 300\,K and $P$\,=\,0\,GPa,
using the  VASP code. Fig.\ref{FIG:Gr}  shows the results in  comparison with
the experimentally extracted $G(R)$.~\cite{Dias2011} Although differences are
not  dramatic, the  HP1 pair  correlations  appear to  agree with  experiment
somewhat better than those for  $\beta$-cristobalite. As a side result, these
simulations  also indicated  a  high  level of  stability  and robustness  of
HP1-CS$_2$ against thermal  fluctuations. It is believed  that this stability
will be important for later tribochemical studies which we are planning.

\section{Phonons, Raman, and Infrared Absorption Spectra}
\label{SEC:PHONONS}

\begin{figure}[t!]
\centering
\includegraphics[width=1.0\linewidth]{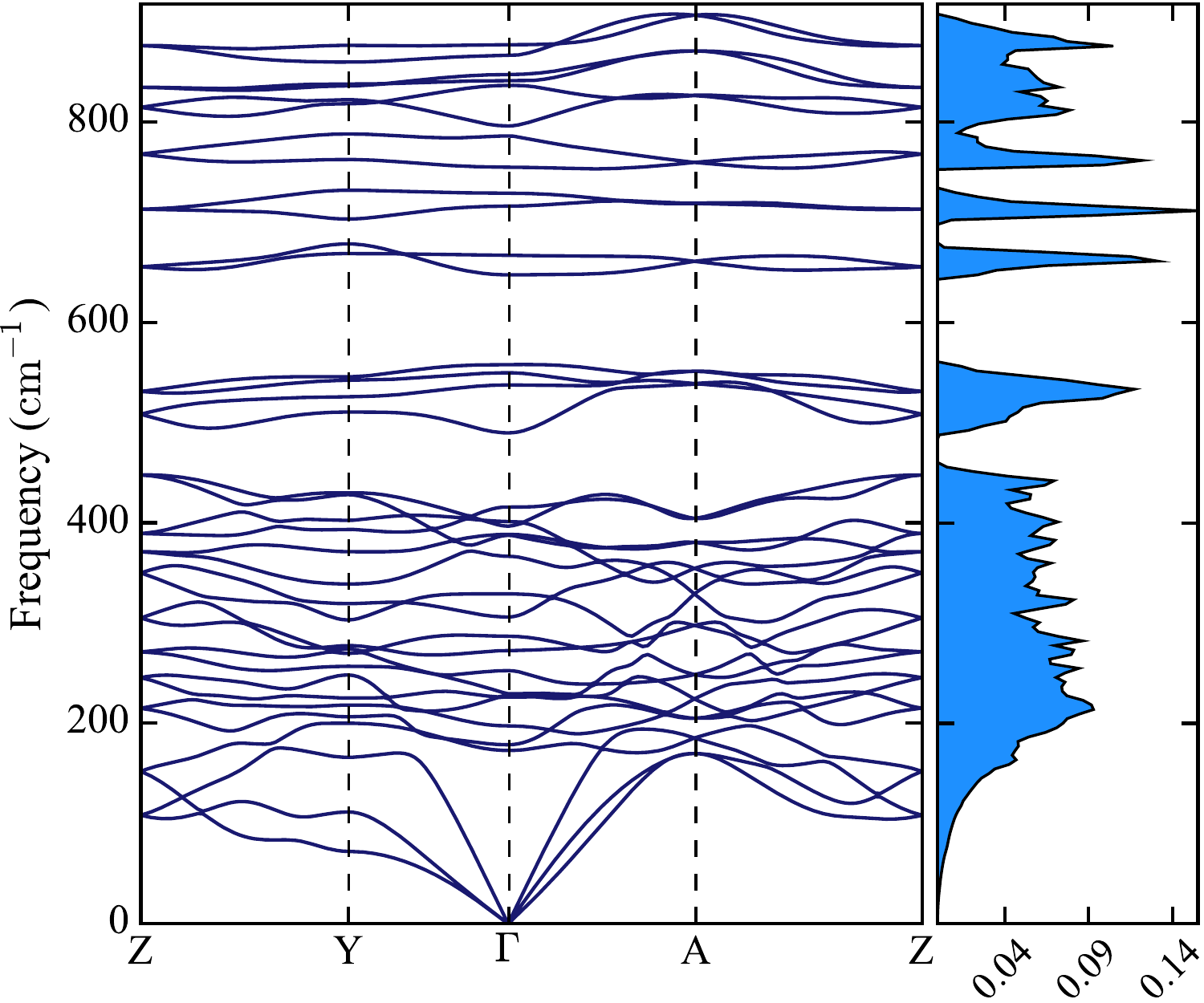}
\caption{(Color online) Calculated phonon  spectrum of the $P2_{1}/c$ layered
structure of CS$_{2}$ at 50\,GPa.}
\label{FIG:4}
\end{figure}

In order to ascertain mechanical  and dynamical stability, and in preparation
for spectroscopy,  we calculated the  phonon spectrum  of the main  HP1 phase
that dominates the phase diagram of CS$_2$  for a wide range of pressure.  As
shown in  Fig.~\ref{FIG:4}, at  P=50 GPa  all mode  frequencies are  real and
positive, confirming  the mechanical stability of  the structure.  Comparison
with calculated phonons of the $\beta$-cristobalite structure~\cite{Dias2011}
shows that modes of the HP1  layered structure are slightly stiffer, although
there is a  fair amount of overall similarity.  Phonon  calculations for HP2,
prohibitively  expensive because  of the  large 36-atom  unit cell,  were not
attempted, also given the uncertain stability of this phase.

\begin{figure}[t!]
\centering
\includegraphics[width=0.99\linewidth]{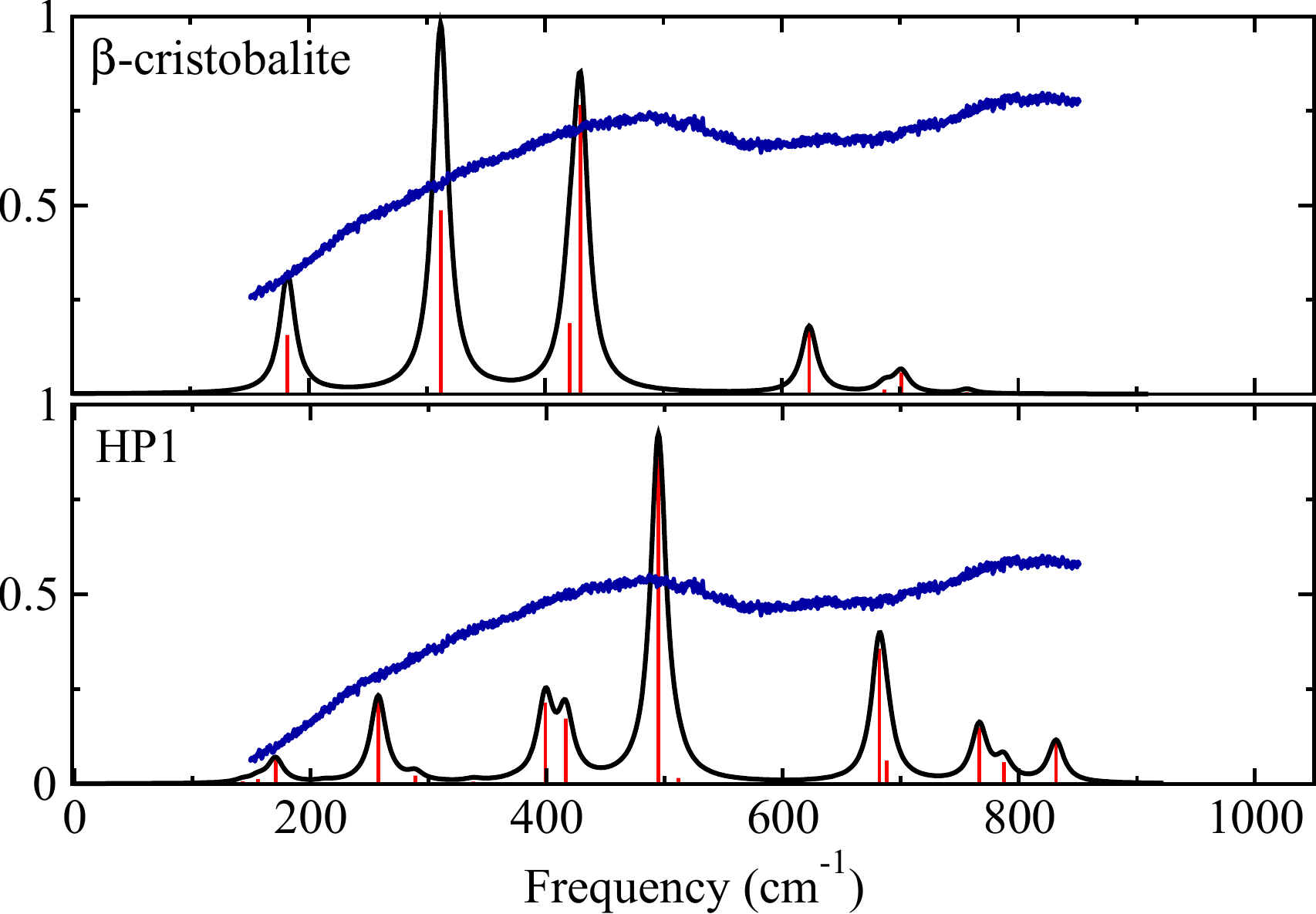}
\caption{(Color online) Predicted Raman spectra of two competing
  CS$_2$ structures at 50 GPa, compared with measurements at 50\,GPa,
  297\,K.\cite{Dias2011} The high frequency secondary peak near 800\,cm$^{-1}$ 
  is only present in layered $P2_{1}/c$ and absent in
  $\beta$-cristobalite. Also the low frequency spectrum is better
  reproduced by HP1 than by $\beta$-cristobalite.}
\label{FIG:5}
\end{figure}

Based on these phonon calculations,  we subsequently calculated the Raman and
IR   absorption   spectra  of   the   HP1   structure  in   comparison   with
$\beta$-cristobalite.  These  spectral  calculations  require  an  insulating
electronic structure. Therefore, even if  available data are mostly at higher
pressures, we conducted  our spectral calculations at 20\,GPa  where both HP1
and  $\beta$-cristobalite   still  have   insulating  LDA   band  structures.
Actually, we found  that LDA at 20\,GPa  yielded a similar volume  to that of
GGA at 30\,GPa, and that the crystal structures underlying these LDA spectral
calculations  are not  very  different  from those  of  our previous  50\,GPa
structural  determinations.  As  shown  in  Fig.\ref{FIG:5}  the  differences
between the layered HP1 and $\beta$-cristobalite predicted spectra are major.
The  Raman spectrum  of  $\beta$-cristobalite  has a  main  (twin) peak  near
300\,cm$^{-1}$, a  second main peak  near 400\,cm$^-1$, much  weaker features
near 600--700\,cm$^-1$,  and nothing  at higher  frequencies.  The  HP1 Raman
predicted spectrum  exhibits instead a  much larger peak  near 500\,cm$^{-1}$
and considerable spectral intensity at 700 and also 800\,cm$^{-1}$.  Both the
500  and the  800 features  agree much  better with  experimental Raman  data
(50\,GPa,  297\,K).~\cite{Dias2011}  This  proves  that HP1  is the  dominant
phase of CS$_2$, as opposed to $\beta$-cristobalite, near 50\,GPa.

\begin{figure}[t!]
\centering
\includegraphics[width=0.99\linewidth]{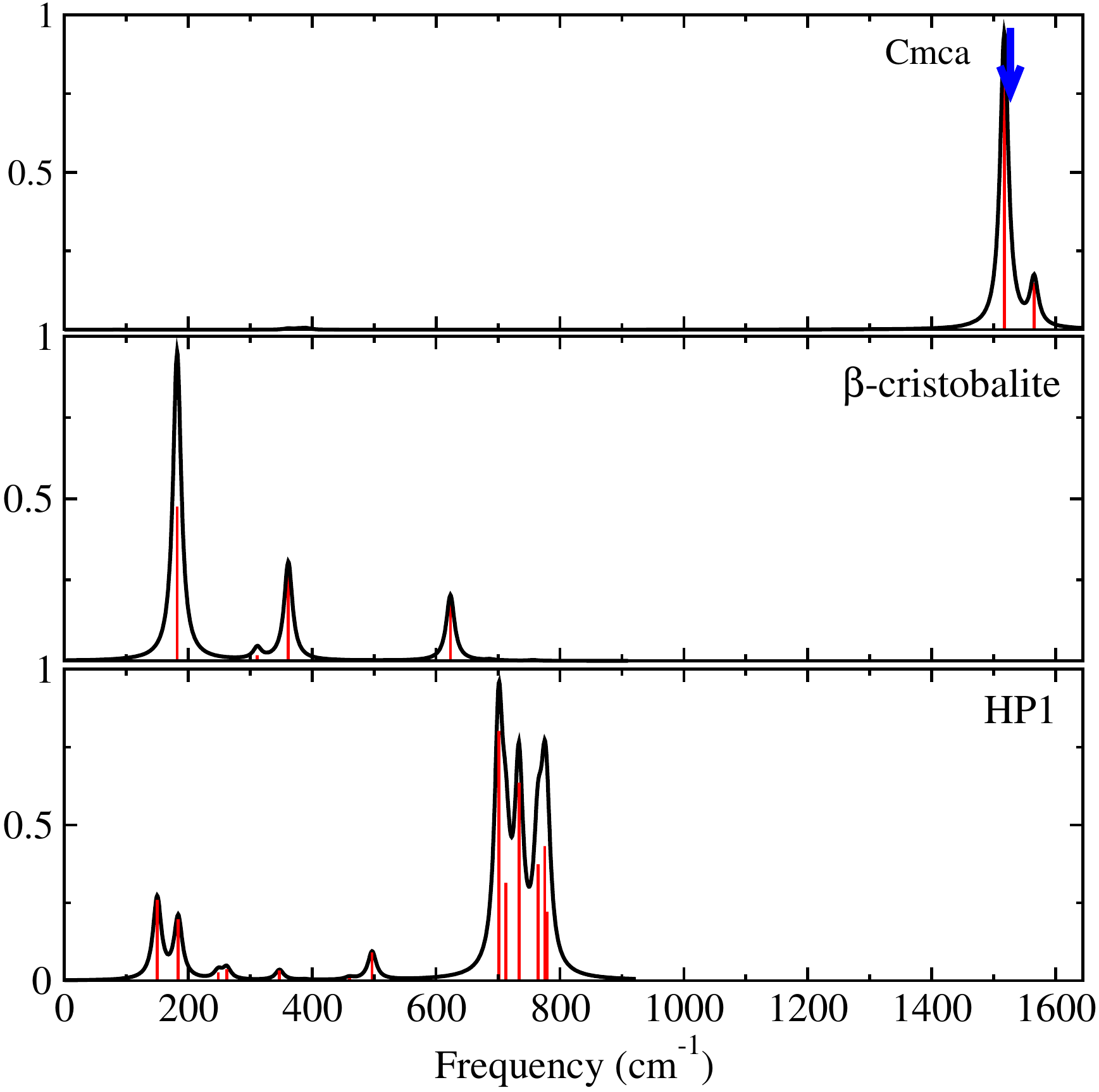}
\caption{(Color online) Calculated IR spectra of different
  CS$_{2}$ structures.  Note the stiffer frequencies of layered HP1
  compared with $\beta$-cristobalite. Blue arrows indicate the
  experimental IR peak positions of molecular
  CS$_{2}$.~\cite{Agnew1988,Yamada1963} }
\label{FIG:IR}
\end{figure}

Besides Raman, high pressure systems should also permit the measurement of IR
absorption. We therefore calculated IR spectra, which not surprisingly turned
out to be very different for the layered HP1 and for $\beta$-cristobalite. As
can be observed  in Fig.\ref{FIG:IR}, the main predicted  absorption peaks of
$\beta$-cristobalite are close to $\approx$  200, 350 and 610\,cm$^{-1}$.  In
the  HP1   phase  instead,   after  a  weaker   structure  between   150  and
200\,cm$^{-1}$ there  is a large  and broad  absorption band between  700 and
800\,cm$^{-1}$, a range  where $\beta$-cristobalite should be  IR silent.  In
future data,  this unmistakeable difference  of IR absorption  spectra should
stand clearly  out.  Experimental IR data  exist apparently only for  the low
pressure molecular structure.  The IR  peak positions observed for the $Cmca$
structure represented by arrows in  Fig.\ref{FIG:IR} agree very well with our
calculations.

\section{Electronic Structure}
\label{SEC:ELECTRONIC}

\begin{figure}[t!]
\centering
\includegraphics[width=1.0\linewidth]{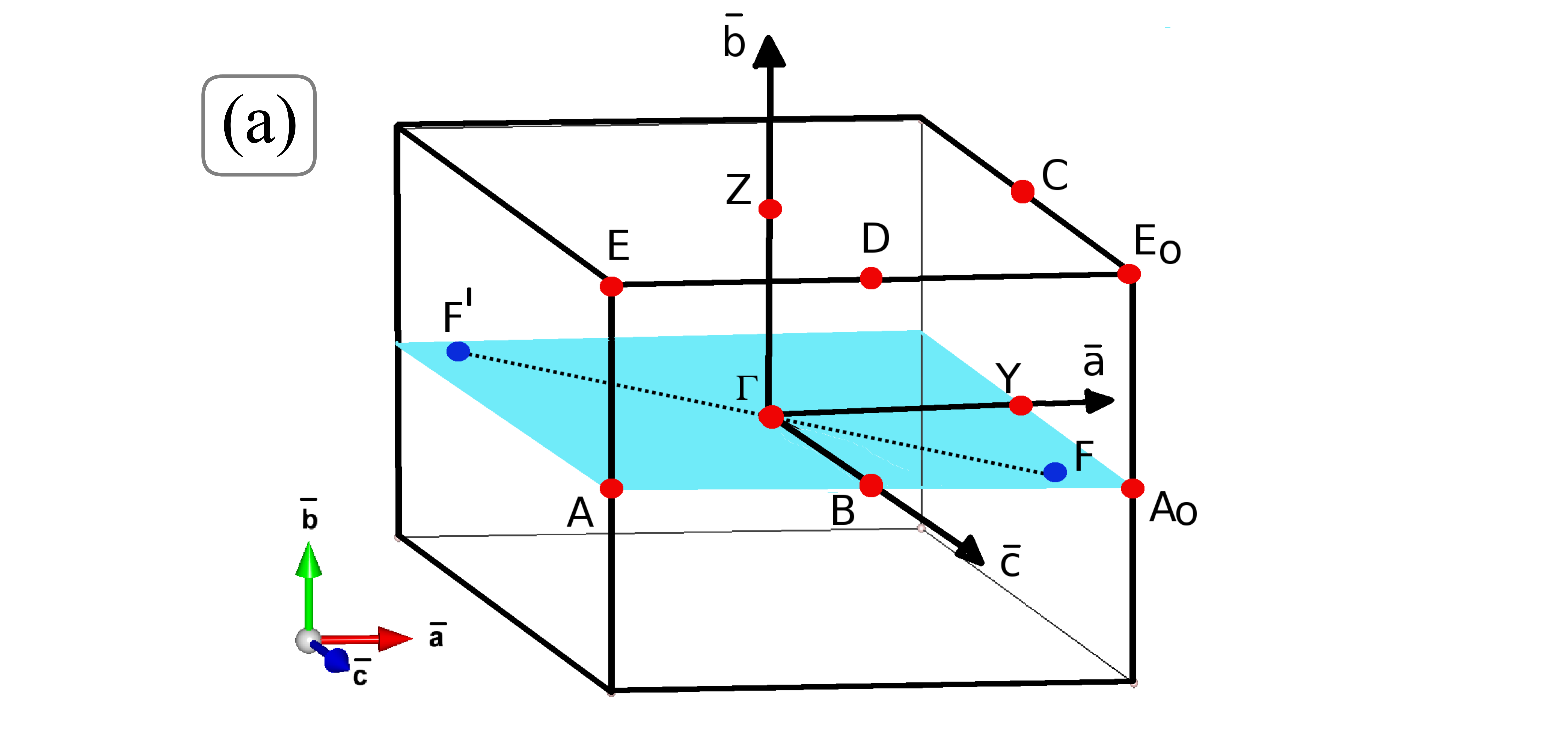}
\includegraphics[width=0.99\linewidth]{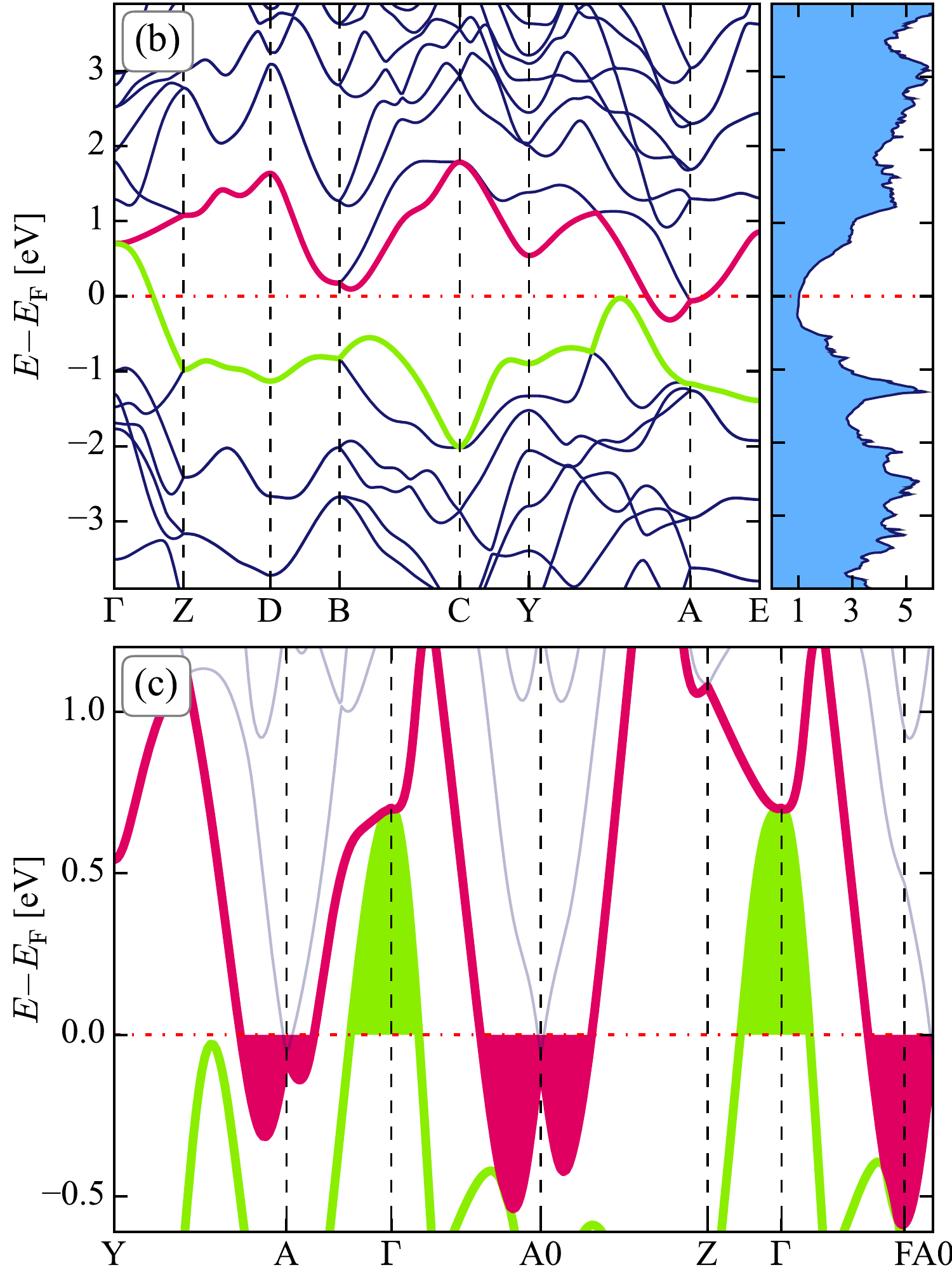}
\caption{(Color  online) (a)  The parallelepiped  Brillouin zone  of the  HP1
$P2_{1}/c$ structure.\cite{bilbao1,bilbao2} F\,=\,(0.3636,  0.0000, 0.4545)
is a  general point in the  Brillouin zone.  (b) PBE  electronic structure of
HP1-CS$_2$ at 40\,GPa,  above the insulator-metal transition.  (c) Bands near
the Fermi  level showing a  single hole pocket  at $\Gamma$ and  two electron
pockets at $F$  and F$^{\prime}$ = $-$F.}
\label{FIG:7}
\end{figure}

DFT calculations yield the electron band structure of all low enthalpy phases
of  CS$_{2}$.   All low  pressure  structures  are insulating.   The  DFT-PBE
electronic   band  structure   of  the   layered  HP1   phase  is   shown  in
Fig.\ref{FIG:7}(a)  at 40\,GPa,  above the  critical metallization  pressure,
where the band gap of HP1-CS$_{2}$ closes. Metallization of HP1-CS$_2$ occurs
at  about  30\,GPa   within  the  PBE  functional,   an  approximation  which
notoriously underestimates the gap  and therefore the metallization pressure.
We repeated the CS$_2$ calculations using B3LYP hybrid functionals, and found
the metallization  pressure to increase  to $\approx$50\,GPa, a value  now in
excellent  agreement with  experiment.~\cite{Dias2011} It  is interesting  to
note that,  owing to  its initially  smaller band  gap than  CO$_2$, CS$_{2}$
metallizes  readily under  pressure after  turning from  a twofold  molecular
state to  a dense fourfold  solid, unlike  CO$_{2}$ which remains  a wide-gap
insulator long after a similar transformation into $\beta$-cristobalite.  One
interesting question  arising at this  point is whether  superconductivity is
predicted in  the metallic high  pressure state of  HP1-CS$_2$. Unfortunately
our  limited resources  and the  large 12-atom  unit cell  prevented us  from
calculating the electron-phonon interaction $\lambda$ and thus estimating the
critical  temperature  T$_c$.   Nonetheless,  a qualitative  answer  to  that
question,  even before  any  detailed calculations,  is  suggested by  direct
inspection of the band structure  of Fig.\ref{FIG:7}(c). Metallization of HP1
takes place by band overlap, with formation of a hole pocket at the $k\!=\!0$
$\Gamma$ point and of a corresponding pair of electron pockets, at $k$-points
$F$ and $F'$ near $A_0$  and $A$---but off the $A-\!{\Gamma}\!-\!A_{0}$ line.
The electron  density of states  of Fig.\ref{FIG:7}(a) calculated  after band
overlap  is  small,  not  suggestive  of  a  large  electron-phonon  coupling
parameter  $\lambda$. As  in other  cases,  our observation  of band  overlap
metallization~\cite{Kohulak2015}  does   not  offer   a  strong   promise  of
superconductivity,  at  least  of  the   standard  BCS  kind.   However,  the
wavevectors  $\Gamma$-F  and $\Gamma$-F$^{\prime}$,  $\pm$\,(0.3636,  0.0000,
0.4545)  are  electron-hole nesting  vectors  of  HP1-CS$_{2}$ near  the  gap
closing pressure around  50\,GPa. It is possible  that charge-density-wave or
spin-density-wave  static modulations  might  appear  with that  periodicity,
possibly also accompanied by some  related superconducting phase.  We are not
presently  in a  position to  inquire quantitatively  into this  possibility,
which would require newer and different approaches.

\section{Conclusions}
\label{SEC:CONCLUSIONS}

We presented a  theoretical study of high pressure solid  phases of CS$_{2}$.
We discarded the obvious possibility  of decomposition by restricting only to
phases  devoid of  C--C and  of S--S  nearest neighbor  bonds, and  aiming at
uncovering  the metastable  phases  of lowest  enthalpy  through an  unbiased
genetic structure search.  Our main result is that,  contrary to expectations
based on similarity with CO$_{2}$ where twofold coordinated $Cmca$ eventually
turns  into  fourfold  coordinated  $\beta$-cristobalite,  high  pressure  in
CS$_{2}$  eventually converts  $Cmca$ into  a different  fourfold coordinated
layered  $P2_{1}/c$  phase built  up  of  edge-sharing pairs  of  tetrahedra.
Recently the very same structure, named HP1, has been experimentally reported
for   SiS$_{2}$    at   much   lower    pressures.~\cite{SiS2_2015}   Another
tetrahedra-based structure,  HP2, is also  stabilized between 8  and 30\,GPa,
but only by  a small enthalpy difference with respect  to HP1 comparable with
our calculation errors.

The proposed  HP1 structure represents  a more plausible  candidate structure
for high  pressure, fourfold  coordinated CS$_{2}$  than those  considered so
far.~\cite{Dias2011,  Dias2013} Both  calculated pair  correlations and  Raman
spectra    agree    better    with    existing    data    than    those    of
$\beta$-cristobalite. It will be a  challenge for the experimentalists to try
to  prepare this  phase  in a  crystalline state  amenable  to more  accurate
investigation in  the future. To  that end we provide  substantial additional
predictions,  in  particular  of  IR  spectra,  that  should  be  crucial  in
identifying the new phase.
Despite  its  intrinsic  metastability, the  HP1-CS$_{2}$  structure  appears
exceptionally robust. These  qualities make CS$_{2}$ a  good candidate system
for studies  of high  pressure simulated tribochemistry,  a project  which is
presently going on in the Trieste group.

Even though  the layered HP1  structure is  metallic above 50\,GPa,  it seems
unlikely   that   it  should   account   for   the  experimentally   observed
superconductivity,~\cite{Dias2013}  because   the  density  of   states,  and
therefore  the dimensionless  $\lambda$, is  likely to  remain low  after the
band-overlap metallic  state.  Although  a reasonable  hypothesis could  be a
possible partial decomposition of CS$_{2}$, with creation of some free sulfur
filaments or other  non stoichiometric products, we are not  in a position to
address that occurrence here.

We note  in closing that  at pressures  just before metallization,  where the
band gap  closing of insulating  HP1-CS$_{2}$ is indirect, the  crystal might
develop   a   narrow    charge-density-wave   or   spin-density-wave   phase,
characterized     by    a     nesting    wavevector     close    to     $\pm$
F.~\cite{Jerome1967} Although there  has been so far no  observation of this
kind in high pressure experiments,  this possibility, which we also suggested
for MoS$_2$,~\cite{Hromadova2013} seems worth pursuing.

Last and perhaps most  important, our work provides a new  link into the high
pressure crystal chemistry of the archetypal family of IV-VI AB$_2$ compounds
made of light elements where  previously only some cross similarities between
CO$_{2}$, SiO$_{2}$,  and SiS$_{2}$  were discussed.  We  show that  there is
some structural  kinship of CS$_{2}$ not  just to CO$_{2}$ at  low pressures,
but eventually also to SiS$_{2}$ at higher pressures.

\begin{acknowledgments} Work  in Trieste  was supported  by the  ERC Advanced
Grant  320796 --  MODPHYSFRICT  which  also covered  the  position  of YC  in
Trieste; by the European Union FP7-NMP-2011-EU-Japan project LEMSUPER; and by
PRIN/COFIN  Contract PRIN-2010LLKJBX\_004.   We acknowledge  discussions with
R.~Hlubina and  P.~Mach.  A CINECA  award 2013 of high  performance computing
resources was  instrumental to the  realization of this  work. Work by  RM in
Bratislava was supported by the  Slovak Research and Development Agency grant
No.~APVV-0108-11 and VEGA grant No.~1/0904/15.
\end{acknowledgments}

\begin{thebibliography}{34}%
\makeatletter
\providecommand \@ifxundefined [1]{%
 \@ifx{#1\undefined}
}%
\providecommand \@ifnum [1]{%
 \ifnum #1\expandafter \@firstoftwo
 \else \expandafter \@secondoftwo
 \fi
}%
\providecommand \@ifx [1]{%
 \ifx #1\expandafter \@firstoftwo
 \else \expandafter \@secondoftwo
 \fi
}%
\providecommand \natexlab [1]{#1}%
\providecommand \enquote  [1]{``#1''}%
\providecommand \bibnamefont  [1]{#1}%
\providecommand \bibfnamefont [1]{#1}%
\providecommand \citenamefont [1]{#1}%
\providecommand \href@noop [0]{\@secondoftwo}%
\providecommand \href [0]{\begingroup \@sanitize@url \@href}%
\providecommand \@href[1]{\@@startlink{#1}\@@href}%
\providecommand \@@href[1]{\endgroup#1\@@endlink}%
\providecommand \@sanitize@url [0]{\catcode `\\12\catcode `\$12\catcode
  `\&12\catcode `\#12\catcode `\^12\catcode `\_12\catcode `\%12\relax}%
\providecommand \@@startlink[1]{}%
\providecommand \@@endlink[0]{}%
\providecommand \url  [0]{\begingroup\@sanitize@url \@url }%
\providecommand \@url [1]{\endgroup\@href {#1}{\urlprefix }}%
\providecommand \urlprefix  [0]{URL }%
\providecommand \Eprint [0]{\href }%
\providecommand \doibase [0]{http://dx.doi.org/}%
\providecommand \selectlanguage [0]{\@gobble}%
\providecommand \bibinfo  [0]{\@secondoftwo}%
\providecommand \bibfield  [0]{\@secondoftwo}%
\providecommand \translation [1]{[#1]}%
\providecommand \BibitemOpen [0]{}%
\providecommand \bibitemStop [0]{}%
\providecommand \bibitemNoStop [0]{.\EOS\space}%
\providecommand \EOS [0]{\spacefactor3000\relax}%
\providecommand \BibitemShut  [1]{\csname bibitem#1\endcsname}%
\let\auto@bib@innerbib\@empty
\bibitem [{\citenamefont {Iota}\ \emph {et~al.}(1999)\citenamefont {Iota},
  \citenamefont {Yoo},\ and\ \citenamefont {Cynn}}]{Iota1999}%
  \BibitemOpen
  \bibfield  {author} {\bibinfo {author} {\bibfnamefont {V.}~\bibnamefont
  {Iota}}, \bibinfo {author} {\bibfnamefont {C.~S.}\ \bibnamefont {Yoo}}, \
  and\ \bibinfo {author} {\bibfnamefont {H.}~\bibnamefont {Cynn}},\ }\href
  {\doibase 10.1126/science.283.5407.1510} {\bibfield  {journal} {\bibinfo
  {journal} {Science}\ }\textbf {\bibinfo {volume} {283}},\ \bibinfo {pages}
  {1510} (\bibinfo {year} {1999})}\BibitemShut {NoStop}%
\bibitem [{\citenamefont {Serra}\ \emph {et~al.}(1999)\citenamefont {Serra},
  \citenamefont {Cavazzoni}, \citenamefont {Chiarotti}, \citenamefont
  {Scandolo},\ and\ \citenamefont {Tosatti}}]{Serra1999}%
  \BibitemOpen
  \bibfield  {author} {\bibinfo {author} {\bibfnamefont {S.}~\bibnamefont
  {Serra}}, \bibinfo {author} {\bibfnamefont {C.}~\bibnamefont {Cavazzoni}},
  \bibinfo {author} {\bibfnamefont {G.~L.}\ \bibnamefont {Chiarotti}}, \bibinfo
  {author} {\bibfnamefont {S.}~\bibnamefont {Scandolo}}, \ and\ \bibinfo
  {author} {\bibfnamefont {E.}~\bibnamefont {Tosatti}},\ }\href {\doibase
  10.1126/science.284.5415.788} {\bibfield  {journal} {\bibinfo  {journal}
  {Science}\ }\textbf {\bibinfo {volume} {284}},\ \bibinfo {pages} {788}
  (\bibinfo {year} {1999})}\BibitemShut {NoStop}%
\bibitem [{\citenamefont {Santoro}\ \emph {et~al.}(2012)\citenamefont
  {Santoro}, \citenamefont {Gorelli}, \citenamefont {Bini}, \citenamefont
  {Haines}, \citenamefont {Cambon}, \citenamefont {Levelut}, \citenamefont
  {Montoya},\ and\ \citenamefont {Scandolo}}]{Santoro2012}%
  \BibitemOpen
  \bibfield  {author} {\bibinfo {author} {\bibfnamefont {M.}~\bibnamefont
  {Santoro}}, \bibinfo {author} {\bibfnamefont {F.~A.}\ \bibnamefont
  {Gorelli}}, \bibinfo {author} {\bibfnamefont {R.}~\bibnamefont {Bini}},
  \bibinfo {author} {\bibfnamefont {J.}~\bibnamefont {Haines}}, \bibinfo
  {author} {\bibfnamefont {O.}~\bibnamefont {Cambon}}, \bibinfo {author}
  {\bibfnamefont {C.}~\bibnamefont {Levelut}}, \bibinfo {author} {\bibfnamefont
  {J.~A.}\ \bibnamefont {Montoya}}, \ and\ \bibinfo {author} {\bibfnamefont
  {S.}~\bibnamefont {Scandolo}},\ }\href {\doibase 10.1073/pnas.1118791109}
  {\bibfield  {journal} {\bibinfo  {journal} {Proc.~Natl.~Acad.~Sci.}\ }\textbf
  {\bibinfo {volume} {109}},\ \bibinfo {pages} {5176} (\bibinfo {year}
  {2012})}\BibitemShut {NoStop}%
\bibitem [{\citenamefont {Sun}\ \emph {et~al.}(2009)\citenamefont {Sun},
  \citenamefont {Klug}, \citenamefont {Marto\^{n}\'{a}k}, \citenamefont
  {Montoya}, \citenamefont {Lee}, \citenamefont {Scandolo},\ and\ \citenamefont
  {Tosatti}}]{sun2009}%
  \BibitemOpen
  \bibfield  {author} {\bibinfo {author} {\bibfnamefont {J.}~\bibnamefont
  {Sun}}, \bibinfo {author} {\bibfnamefont {D.~D.}\ \bibnamefont {Klug}},
  \bibinfo {author} {\bibfnamefont {R.}~\bibnamefont {Marto\^{n}\'{a}k}},
  \bibinfo {author} {\bibfnamefont {J.~A.}\ \bibnamefont {Montoya}}, \bibinfo
  {author} {\bibfnamefont {M.-S.}\ \bibnamefont {Lee}}, \bibinfo {author}
  {\bibfnamefont {S.}~\bibnamefont {Scandolo}}, \ and\ \bibinfo {author}
  {\bibfnamefont {E.}~\bibnamefont {Tosatti}},\ }\href {\doibase
  10.1073/pnas.0812624106} {\bibfield  {journal} {\bibinfo  {journal}
  {Proc.~Natl.~Acad.~Sci.}\ }\textbf {\bibinfo {volume} {106}},\ \bibinfo
  {pages} {6077} (\bibinfo {year} {2009})}\BibitemShut {NoStop}%
\bibitem [{\citenamefont {Cox}\ \emph {et~al.}(1984)\citenamefont {Cox},
  \citenamefont {Wagman},\ and\ \citenamefont {Medvedev}}]{CO2-Hf_1}%
  \BibitemOpen
  \bibfield  {author} {\bibinfo {author} {\bibfnamefont {J.~D.}\ \bibnamefont
  {Cox}}, \bibinfo {author} {\bibfnamefont {D.~D.}\ \bibnamefont {Wagman}}, \
  and\ \bibinfo {author} {\bibfnamefont {V.~A.}\ \bibnamefont {Medvedev}},\
  }\href@noop {} {\emph {\bibinfo {title} {CODATA Key Values for
  Thermodynamics}}}\ (\bibinfo  {publisher} {Hemisphere Publishing Corp., New
  York},\ \bibinfo {year} {1984})\BibitemShut {NoStop}%
\bibitem [{\citenamefont {Chase}(1998)}]{Hf}%
  \BibitemOpen
  \bibfield  {author} {\bibinfo {author} {\bibfnamefont {M.~J.}\ \bibnamefont
  {Chase}},\ }\href@noop {} {\bibfield  {journal} {\bibinfo  {journal} {J.
  Phys. Chem. Ref. Data, Monograph}\ }\textbf {\bibinfo {volume} {9}},\
  \bibinfo {pages} {1} (\bibinfo {year} {1998})}\BibitemShut {NoStop}%
\bibitem [{\citenamefont {Dias}\ \emph {et~al.}(2011)\citenamefont {Dias},
  \citenamefont {Yoo}, \citenamefont {Kim},\ and\ \citenamefont
  {Tse}}]{Dias2011}%
  \BibitemOpen
  \bibfield  {author} {\bibinfo {author} {\bibfnamefont {R.~P.}\ \bibnamefont
  {Dias}}, \bibinfo {author} {\bibfnamefont {C.-S.}\ \bibnamefont {Yoo}},
  \bibinfo {author} {\bibfnamefont {M.}~\bibnamefont {Kim}}, \ and\ \bibinfo
  {author} {\bibfnamefont {J.~S.}\ \bibnamefont {Tse}},\ }\href {\doibase
  10.1103/PhysRevB.84.144104} {\bibfield  {journal} {\bibinfo  {journal}
  {Phys.~Rev.~B}\ }\textbf {\bibinfo {volume} {84}},\ \bibinfo {pages} {144104}
  (\bibinfo {year} {2011})}\BibitemShut {NoStop}%
\bibitem [{\citenamefont {Bolduan}\ \emph {et~al.}(1986)\citenamefont
  {Bolduan}, \citenamefont {Hochheimer},\ and\ \citenamefont
  {Jodl}}]{Bolduan1986}%
  \BibitemOpen
  \bibfield  {author} {\bibinfo {author} {\bibfnamefont {F.}~\bibnamefont
  {Bolduan}}, \bibinfo {author} {\bibfnamefont {H.~D.}\ \bibnamefont
  {Hochheimer}}, \ and\ \bibinfo {author} {\bibfnamefont {H.~J.}\ \bibnamefont
  {Jodl}},\ }\href {\doibase 10.1063/1.450817} {\bibfield  {journal} {\bibinfo
  {journal} {J.~Chem.~Phys.}\ }\textbf {\bibinfo {volume} {84}},\ \bibinfo
  {pages} {6997} (\bibinfo {year} {1986})}\BibitemShut {NoStop}%
\bibitem [{\citenamefont {Dias}\ \emph {et~al.}(2013)\citenamefont {Dias},
  \citenamefont {Yoo}, \citenamefont {Struzhkin}, \citenamefont {Kim},
  \citenamefont {Muramatsu}, \citenamefont {Matsuoka}, \citenamefont {Ohishi},\
  and\ \citenamefont {Sinogeikin}}]{Dias2013}%
  \BibitemOpen
  \bibfield  {author} {\bibinfo {author} {\bibfnamefont {R.~P.}\ \bibnamefont
  {Dias}}, \bibinfo {author} {\bibfnamefont {C.-S.}\ \bibnamefont {Yoo}},
  \bibinfo {author} {\bibfnamefont {V.~V.}\ \bibnamefont {Struzhkin}}, \bibinfo
  {author} {\bibfnamefont {M.}~\bibnamefont {Kim}}, \bibinfo {author}
  {\bibfnamefont {T.}~\bibnamefont {Muramatsu}}, \bibinfo {author}
  {\bibfnamefont {T.}~\bibnamefont {Matsuoka}}, \bibinfo {author}
  {\bibfnamefont {Y.}~\bibnamefont {Ohishi}}, \ and\ \bibinfo {author}
  {\bibfnamefont {S.}~\bibnamefont {Sinogeikin}},\ }\href {\doibase
  10.1073/pnas.1305129110} {\bibfield  {journal} {\bibinfo  {journal}
  {Proc.~Natl.~Acad.~Sci.}\ }\textbf {\bibinfo {volume} {29}},\ \bibinfo
  {pages} {11720} (\bibinfo {year} {2013})}\BibitemShut {NoStop}%
\bibitem [{\citenamefont {Evers}\ \emph {et~al.}(2015)\citenamefont {Evers},
  \citenamefont {Mayer}, \citenamefont {M{\"o}ckl}, \citenamefont {Oehlinger},
  \citenamefont {K{\"o}ppe},\ and\ \citenamefont {Schn{\"o}ckel}}]{SiS2_2015}%
  \BibitemOpen
  \bibfield  {author} {\bibinfo {author} {\bibfnamefont {J.}~\bibnamefont
  {Evers}}, \bibinfo {author} {\bibfnamefont {P.}~\bibnamefont {Mayer}},
  \bibinfo {author} {\bibfnamefont {L.}~\bibnamefont {M{\"o}ckl}}, \bibinfo
  {author} {\bibfnamefont {G.}~\bibnamefont {Oehlinger}}, \bibinfo {author}
  {\bibfnamefont {R.}~\bibnamefont {K{\"o}ppe}}, \ and\ \bibinfo {author}
  {\bibfnamefont {H.}~\bibnamefont {Schn{\"o}ckel}},\ }\href {\doibase
  10.1021/ic501825r} {\bibfield  {journal} {\bibinfo  {journal} {Inorg. Chem.}\
  }\textbf {\bibinfo {volume} {54}},\ \bibinfo {pages} {1240} (\bibinfo {year}
  {2015})}\BibitemShut {NoStop}%
\bibitem [{\citenamefont {Glass}\ \emph {et~al.}(2006)\citenamefont {Glass},
  \citenamefont {Oganov},\ and\ \citenamefont {Hansen}}]{USPEX1}%
  \BibitemOpen
  \bibfield  {author} {\bibinfo {author} {\bibfnamefont {C.~W.}\ \bibnamefont
  {Glass}}, \bibinfo {author} {\bibfnamefont {A.~R.}\ \bibnamefont {Oganov}}, \
  and\ \bibinfo {author} {\bibfnamefont {N.}~\bibnamefont {Hansen}},\ }\href
  {\doibase 10.1016/j.cpc.2006.07.020} {\bibfield  {journal} {\bibinfo
  {journal} {Comput.~Phys.~Commun.}\ }\textbf {\bibinfo {volume} {175}},\
  \bibinfo {pages} {713} (\bibinfo {year} {2006})}\BibitemShut {NoStop}%
\bibitem [{\citenamefont {Oganov}\ and\ \citenamefont {Glass}(2006)}]{USPEX2}%
  \BibitemOpen
  \bibfield  {author} {\bibinfo {author} {\bibfnamefont {A.~R.}\ \bibnamefont
  {Oganov}}\ and\ \bibinfo {author} {\bibfnamefont {C.~W.}\ \bibnamefont
  {Glass}},\ }\href {\doibase 10.1063/1.2210932} {\bibfield  {journal}
  {\bibinfo  {journal} {J.~Chem.~Phys.}\ }\textbf {\bibinfo {volume} {124}},\
  \bibinfo {pages} {244704} (\bibinfo {year} {2006})}\BibitemShut {NoStop}%
\bibitem [{\citenamefont {Lyakhov}\ \emph {et~al.}(2013)\citenamefont
  {Lyakhov}, \citenamefont {Oganov}, \citenamefont {Stokes},\ and\
  \citenamefont {Zhu}}]{USPEX3}%
  \BibitemOpen
  \bibfield  {author} {\bibinfo {author} {\bibfnamefont {A.~O.}\ \bibnamefont
  {Lyakhov}}, \bibinfo {author} {\bibfnamefont {A.~R.}\ \bibnamefont {Oganov}},
  \bibinfo {author} {\bibfnamefont {H.~T.}\ \bibnamefont {Stokes}}, \ and\
  \bibinfo {author} {\bibfnamefont {Q.}~\bibnamefont {Zhu}},\ }\href {\doibase
  10.1016/j.cpc.2012.12.009} {\bibfield  {journal} {\bibinfo  {journal}
  {Comput. Phys. Commun.}\ }\textbf {\bibinfo {volume} {184}},\ \bibinfo
  {pages} {1172} (\bibinfo {year} {2013})}\BibitemShut {NoStop}%
\bibitem [{\citenamefont {Oganov}\ \emph {et~al.}(2011)\citenamefont {Oganov},
  \citenamefont {Lyakhov},\ and\ \citenamefont {Valle}}]{USPEX4}%
  \BibitemOpen
  \bibfield  {author} {\bibinfo {author} {\bibfnamefont {A.~R.}\ \bibnamefont
  {Oganov}}, \bibinfo {author} {\bibfnamefont {A.~O.}\ \bibnamefont {Lyakhov}},
  \ and\ \bibinfo {author} {\bibfnamefont {M.}~\bibnamefont {Valle}},\ }\href
  {\doibase 10.1021/ar1001318} {\bibfield  {journal} {\bibinfo  {journal}
  {Acc.~Chem.~Res.}\ }\textbf {\bibinfo {volume} {44}},\ \bibinfo {pages} {227}
  (\bibinfo {year} {2011})}\BibitemShut {NoStop}%
\bibitem [{\citenamefont {Zhu}\ \emph {et~al.}(2012)\citenamefont {Zhu},
  \citenamefont {Oganov}, \citenamefont {Glass},\ and\ \citenamefont
  {Stokes}}]{USPEX5M}%
  \BibitemOpen
  \bibfield  {author} {\bibinfo {author} {\bibfnamefont {Q.}~\bibnamefont
  {Zhu}}, \bibinfo {author} {\bibfnamefont {A.~R.}\ \bibnamefont {Oganov}},
  \bibinfo {author} {\bibfnamefont {C.~W.}\ \bibnamefont {Glass}}, \ and\
  \bibinfo {author} {\bibfnamefont {H.~T.}\ \bibnamefont {Stokes}},\ }\href
  {\doibase 10.1107/S0108768112017466} {\bibfield  {journal} {\bibinfo
  {journal} {Acta Crystallogr. Sect. B}\ }\textbf {\bibinfo {volume} {68}},\
  \bibinfo {pages} {215} (\bibinfo {year} {2012})}\BibitemShut {NoStop}%
\bibitem [{\citenamefont {Perdew}\ \emph {et~al.}(1997)\citenamefont {Perdew},
  \citenamefont {Burke},\ and\ \citenamefont {Ernzerhof}}]{PBE}%
  \BibitemOpen
  \bibfield  {author} {\bibinfo {author} {\bibfnamefont {J.~P.}\ \bibnamefont
  {Perdew}}, \bibinfo {author} {\bibfnamefont {K.}~\bibnamefont {Burke}}, \
  and\ \bibinfo {author} {\bibfnamefont {M.}~\bibnamefont {Ernzerhof}},\ }\href
  {\doibase 10.1103/PhysRevLett.78.1396} {\bibfield  {journal} {\bibinfo
  {journal} {Phys.~Rev.~Lett.}\ }\textbf {\bibinfo {volume} {78}},\ \bibinfo
  {pages} {1396} (\bibinfo {year} {1997})}\BibitemShut {NoStop}%
\bibitem [{\citenamefont {Kresse}\ and\ \citenamefont
  {Furthm\"uller}(1996)}]{VASP1}%
  \BibitemOpen
  \bibfield  {author} {\bibinfo {author} {\bibfnamefont {G.}~\bibnamefont
  {Kresse}}\ and\ \bibinfo {author} {\bibfnamefont {J.}~\bibnamefont
  {Furthm\"uller}},\ }\href {\doibase 10.1103/PhysRevB.54.11169} {\bibfield
  {journal} {\bibinfo  {journal} {Phys. Rev. B}\ }\textbf {\bibinfo {volume}
  {54}},\ \bibinfo {pages} {11169} (\bibinfo {year} {1996})}\BibitemShut
  {NoStop}%
\bibitem [{\citenamefont {Bl\"ochl}(1994)}]{PAW1}%
  \BibitemOpen
  \bibfield  {author} {\bibinfo {author} {\bibfnamefont {P.~E.}\ \bibnamefont
  {Bl\"ochl}},\ }\href {\doibase 10.1103/PhysRevB.50.17953} {\bibfield
  {journal} {\bibinfo  {journal} {Phys. Rev. B}\ }\textbf {\bibinfo {volume}
  {50}},\ \bibinfo {pages} {17953} (\bibinfo {year} {1994})}\BibitemShut
  {NoStop}%
\bibitem [{\citenamefont {Kresse}\ and\ \citenamefont {Joubert}(1999)}]{PAW2}%
  \BibitemOpen
  \bibfield  {author} {\bibinfo {author} {\bibfnamefont {G.}~\bibnamefont
  {Kresse}}\ and\ \bibinfo {author} {\bibfnamefont {D.}~\bibnamefont
  {Joubert}},\ }\href {\doibase 10.1103/PhysRevB.59.1758} {\bibfield  {journal}
  {\bibinfo  {journal} {Phys. Rev. B}\ }\textbf {\bibinfo {volume} {59}},\
  \bibinfo {pages} {1758} (\bibinfo {year} {1999})}\BibitemShut {NoStop}%
\bibitem [{\citenamefont {Klime\v{s}}\ \emph {et~al.}(2010)\citenamefont
  {Klime\v{s}}, \citenamefont {Bowler},\ and\ \citenamefont
  {Michaelides}}]{Klimes2011}%
  \BibitemOpen
  \bibfield  {author} {\bibinfo {author} {\bibfnamefont {J.}~\bibnamefont
  {Klime\v{s}}}, \bibinfo {author} {\bibfnamefont {D.~R.}\ \bibnamefont
  {Bowler}}, \ and\ \bibinfo {author} {\bibfnamefont {A.}~\bibnamefont
  {Michaelides}},\ }\href {\doibase 10.1088/0953-8984/22/2/022201} {\bibfield
  {journal} {\bibinfo  {journal} {J.~Phys.~Condens.~Matter}\ }\textbf {\bibinfo
  {volume} {22}},\ \bibinfo {pages} {022201} (\bibinfo {year}
  {2010})}\BibitemShut {NoStop}%
\bibitem [{\citenamefont {Klime\v{s}}\ \emph {et~al.}(2011)\citenamefont
  {Klime\v{s}}, \citenamefont {Bowler},\ and\ \citenamefont
  {Michaelides}}]{Klimes2011_2}%
  \BibitemOpen
  \bibfield  {author} {\bibinfo {author} {\bibfnamefont {J.}~\bibnamefont
  {Klime\v{s}}}, \bibinfo {author} {\bibfnamefont {D.~R.}\ \bibnamefont
  {Bowler}}, \ and\ \bibinfo {author} {\bibfnamefont {A.}~\bibnamefont
  {Michaelides}},\ }\href {\doibase
  http://dx.doi.org/10.1103/PhysRevB.83.195131} {\bibfield  {journal} {\bibinfo
   {journal} {Phys.~Rev.~B}\ }\textbf {\bibinfo {volume} {83}},\ \bibinfo
  {pages} {195131} (\bibinfo {year} {2011})}\BibitemShut {NoStop}%
\bibitem [{\citenamefont {Dion}\ \emph {et~al.}(2004)\citenamefont {Dion},
  \citenamefont {Rydberg}, \citenamefont {Schr\"oder}, \citenamefont
  {Langreth},\ and\ \citenamefont {Lundqvist}}]{Dion2004}%
  \BibitemOpen
  \bibfield  {author} {\bibinfo {author} {\bibfnamefont {M.}~\bibnamefont
  {Dion}}, \bibinfo {author} {\bibfnamefont {H.}~\bibnamefont {Rydberg}},
  \bibinfo {author} {\bibfnamefont {E.}~\bibnamefont {Schr\"oder}}, \bibinfo
  {author} {\bibfnamefont {D.~C.}\ \bibnamefont {Langreth}}, \ and\ \bibinfo
  {author} {\bibfnamefont {B.~I.}\ \bibnamefont {Lundqvist}},\ }\href {\doibase
  10.1103/PhysRevLett.92.246401} {\bibfield  {journal} {\bibinfo  {journal}
  {Phys. Rev. Lett.}\ }\textbf {\bibinfo {volume} {92}},\ \bibinfo {pages}
  {246401} (\bibinfo {year} {2004})}\BibitemShut {NoStop}%
\bibitem [{\citenamefont {Giannozzi~{\it et al}}(2009)}]{pwscf}%
  \BibitemOpen
  \bibfield  {author} {\bibinfo {author} {\bibfnamefont {P.}~\bibnamefont
  {Giannozzi~{\it et al}}},\ }\href {\doibase 10.1088/0953-8984/21/39/395502}
  {\bibfield  {journal} {\bibinfo  {journal} {J.~Phys.~Condens.~Matter}\
  }\textbf {\bibinfo {volume} {21}},\ \bibinfo {pages} {395502} (\bibinfo
  {year} {2009})}\BibitemShut {NoStop}%
\bibitem [{\citenamefont {Guseva}\ \emph {et~al.}(1991)\citenamefont {Guseva},
  \citenamefont {Burdina},\ and\ \citenamefont {Semenenko}}]{Guseva1991}%
  \BibitemOpen
  \bibfield  {author} {\bibinfo {author} {\bibfnamefont {T.~A.}\ \bibnamefont
  {Guseva}}, \bibinfo {author} {\bibfnamefont {K.~P.}\ \bibnamefont {Burdina}},
  \ and\ \bibinfo {author} {\bibfnamefont {K.~N.}\ \bibnamefont {Semenenko}},\
  }\href@noop {} {\bibfield  {journal} {\bibinfo  {journal} {Khimiya}\ }\textbf
  {\bibinfo {volume} {32}},\ \bibinfo {pages} {85} (\bibinfo {year}
  {1991})}\BibitemShut {NoStop}%
\bibitem [{\citenamefont {Bader}(1990)}]{Bader1990}%
  \BibitemOpen
  \bibfield  {author} {\bibinfo {author} {\bibfnamefont {R.~F.~W.}\
  \bibnamefont {Bader}},\ }\href@noop {} {\emph {\bibinfo {title} {Atoms in
  Molecules: {A} Quantum Theory}}},\ edited by\ \bibinfo {editor}
  {\bibfnamefont {R.~F.~W.}\ \bibnamefont {Bader}}\ (\bibinfo  {publisher}
  {Oxford University Press},\ \bibinfo {year} {1990})\BibitemShut {NoStop}%
\bibitem [{\citenamefont {Henkelman}\ \emph {et~al.}(2006)\citenamefont
  {Henkelman}, \citenamefont {Arnaldsson},\ and\ \citenamefont
  {J{\'o}nsson}}]{Henkelman2006}%
  \BibitemOpen
  \bibfield  {author} {\bibinfo {author} {\bibfnamefont {G.}~\bibnamefont
  {Henkelman}}, \bibinfo {author} {\bibfnamefont {A.}~\bibnamefont
  {Arnaldsson}}, \ and\ \bibinfo {author} {\bibfnamefont {H.}~\bibnamefont
  {J{\'o}nsson}},\ }\href {\doibase 10.1016/j.commatsci.2005.04.010} {\bibfield
   {journal} {\bibinfo  {journal} {Comput. Mater. Sci}\ }\textbf {\bibinfo
  {volume} {36}},\ \bibinfo {pages} {354} (\bibinfo {year} {2006})}\BibitemShut
  {NoStop}%
\bibitem [{\citenamefont {House}(2008)}]{james_book}%
  \BibitemOpen
  \bibfield  {author} {\bibinfo {author} {\bibfnamefont {J.~E.}\ \bibnamefont
  {House}},\ }\href@noop {} {\emph {\bibinfo {title} {Inorganic Chemistry}}},\
  edited by\ \bibinfo {editor} {\bibfnamefont {J.~E.}\ \bibnamefont {House}}\
  (\bibinfo  {publisher} {Elsevier},\ \bibinfo {year} {2008})\BibitemShut
  {NoStop}%
\bibitem [{\citenamefont {Agnew}\ \emph {et~al.}(1988)\citenamefont {Agnew},
  \citenamefont {Mischke},\ and\ \citenamefont {Swanson}}]{Agnew1988}%
  \BibitemOpen
  \bibfield  {author} {\bibinfo {author} {\bibfnamefont {S.~F.}\ \bibnamefont
  {Agnew}}, \bibinfo {author} {\bibfnamefont {R.~E.}\ \bibnamefont {Mischke}},
  \ and\ \bibinfo {author} {\bibfnamefont {B.~I.}\ \bibnamefont {Swanson}},\
  }\href {\doibase 10.1021/j100325a041} {\bibfield  {journal} {\bibinfo
  {journal} {J.~Phys.~Chem.}\ }\textbf {\bibinfo {volume} {92}},\ \bibinfo
  {pages} {4201} (\bibinfo {year} {1988})}\BibitemShut {NoStop}%
\bibitem [{\citenamefont {Yamada}\ and\ \citenamefont
  {Person}(1963)}]{Yamada1963}%
  \BibitemOpen
  \bibfield  {author} {\bibinfo {author} {\bibfnamefont {H.}~\bibnamefont
  {Yamada}}\ and\ \bibinfo {author} {\bibfnamefont {W.~B.}\ \bibnamefont
  {Person}},\ }\href {\doibase 10.1063/1.1725114} {\bibfield  {journal}
  {\bibinfo  {journal} {J.~Chem.~Phys.}\ }\textbf {\bibinfo {volume} {40}},\
  \bibinfo {pages} {309} (\bibinfo {year} {1963})}\BibitemShut {NoStop}%
\bibitem [{\citenamefont {Aroyo}\ \emph
  {et~al.}(2006{\natexlab{a}})\citenamefont {Aroyo}, \citenamefont
  {Perez-Mato}, \citenamefont {Capillas}, \citenamefont {Kroumova},
  \citenamefont {Ivantchev}, \citenamefont {Madariaga}, \citenamefont {Kirov},\
  and\ \citenamefont {Wondratschek}}]{bilbao1}%
  \BibitemOpen
  \bibfield  {author} {\bibinfo {author} {\bibfnamefont {M.~I.}\ \bibnamefont
  {Aroyo}}, \bibinfo {author} {\bibfnamefont {J.~M.}\ \bibnamefont
  {Perez-Mato}}, \bibinfo {author} {\bibfnamefont {C.}~\bibnamefont
  {Capillas}}, \bibinfo {author} {\bibfnamefont {E.}~\bibnamefont {Kroumova}},
  \bibinfo {author} {\bibfnamefont {S.}~\bibnamefont {Ivantchev}}, \bibinfo
  {author} {\bibfnamefont {G.}~\bibnamefont {Madariaga}}, \bibinfo {author}
  {\bibfnamefont {A.}~\bibnamefont {Kirov}}, \ and\ \bibinfo {author}
  {\bibfnamefont {H.}~\bibnamefont {Wondratschek}},\ }\href {\doibase
  http://dx.doi.org/10.1524/zkri.2006.221.1.15} {\bibfield  {journal} {\bibinfo
   {journal} {Z. Krist.}\ }\textbf {\bibinfo {volume} {221}},\ \bibinfo {pages}
  {15} (\bibinfo {year} {2006}{\natexlab{a}})}\BibitemShut {NoStop}%
\bibitem [{\citenamefont {Aroyo}\ \emph
  {et~al.}(2006{\natexlab{b}})\citenamefont {Aroyo}, \citenamefont {Kirov},
  \citenamefont {Capillas}, \citenamefont {Perez-Mato},\ and\ \citenamefont
  {Wondratschek}}]{bilbao2}%
  \BibitemOpen
  \bibfield  {author} {\bibinfo {author} {\bibfnamefont {M.~I.}\ \bibnamefont
  {Aroyo}}, \bibinfo {author} {\bibfnamefont {A.}~\bibnamefont {Kirov}},
  \bibinfo {author} {\bibfnamefont {C.}~\bibnamefont {Capillas}}, \bibinfo
  {author} {\bibfnamefont {J.~M.}\ \bibnamefont {Perez-Mato}}, \ and\ \bibinfo
  {author} {\bibfnamefont {H.}~\bibnamefont {Wondratschek}},\ }\href {\doibase
  10.1107/S0108767305040286} {\bibfield  {journal} {\bibinfo  {journal} {Acta
  Cryst.}\ }\textbf {\bibinfo {volume} {A62}},\ \bibinfo {pages} {115}
  (\bibinfo {year} {2006}{\natexlab{b}})}\BibitemShut {NoStop}%
\bibitem [{\citenamefont {Kohul\'{a}k}\ \emph {et~al.}(2015)\citenamefont
  {Kohul\'{a}k}, \citenamefont {Marto\v{n}\'{a}k},\ and\ \citenamefont
  {Tosatti}}]{Kohulak2015}%
  \BibitemOpen
  \bibfield  {author} {\bibinfo {author} {\bibfnamefont {O.}~\bibnamefont
  {Kohul\'{a}k}}, \bibinfo {author} {\bibfnamefont {R.}~\bibnamefont
  {Marto\v{n}\'{a}k}}, \ and\ \bibinfo {author} {\bibfnamefont
  {E.}~\bibnamefont {Tosatti}},\ }\href {\doibase 10.1103/PhysRevB.91.144113}
  {\bibfield  {journal} {\bibinfo  {journal} {Phys.~Rev.~B}\ }\textbf {\bibinfo
  {volume} {91}},\ \bibinfo {pages} {144113} (\bibinfo {year}
  {2015})}\BibitemShut {NoStop}%
\bibitem [{\citenamefont {J\'erome}\ \emph {et~al.}(1967)\citenamefont
  {J\'erome}, \citenamefont {Rice},\ and\ \citenamefont {Kohn}}]{Jerome1967}%
  \BibitemOpen
  \bibfield  {author} {\bibinfo {author} {\bibfnamefont {D.}~\bibnamefont
  {J\'erome}}, \bibinfo {author} {\bibfnamefont {T.~M.}\ \bibnamefont {Rice}},
  \ and\ \bibinfo {author} {\bibfnamefont {W.}~\bibnamefont {Kohn}},\ }\href
  {\doibase 10.1103/PhysRev.158.462} {\bibfield  {journal} {\bibinfo  {journal}
  {Phys. Rev.}\ }\textbf {\bibinfo {volume} {158}},\ \bibinfo {pages} {462}
  (\bibinfo {year} {1967})}\BibitemShut {NoStop}%
\bibitem [{\citenamefont {Hromadov\'{a}}\ \emph {et~al.}(2013)\citenamefont
  {Hromadov\'{a}}, \citenamefont {Marto\v{n}\'{a}k},\ and\ \citenamefont
  {Tosatti}}]{Hromadova2013}%
  \BibitemOpen
  \bibfield  {author} {\bibinfo {author} {\bibfnamefont {L.}~\bibnamefont
  {Hromadov\'{a}}}, \bibinfo {author} {\bibfnamefont {R.}~\bibnamefont
  {Marto\v{n}\'{a}k}}, \ and\ \bibinfo {author} {\bibfnamefont
  {E.}~\bibnamefont {Tosatti}},\ }\href {\doibase 10.1103/PhysRevB.87.144105}
  {\bibfield  {journal} {\bibinfo  {journal} {Phys.~Rev.~B}\ }\textbf {\bibinfo
  {volume} {87}},\ \bibinfo {pages} {144105} (\bibinfo {year}
  {2013})}\BibitemShut {NoStop}%
\end{thebibliography}
%
\end{document}